\documentclass[12pt,preprint]{aastex}

\begin{document}


\title{The First Direct Measurements of Magnetic Fields on Very 
Low-Mass Stars}


\author{A. Reiners\altaffilmark{1,2,^\star} \and G. Basri}
\affil{Astronomy Department, University of California, Berkeley, CA
  94720 \email{[areiners, basri]@berkeley.edu}}
\altaffiltext{1}{Hamburger Sternwarte, Universit\"at Hamburg,
  Gojenbergsweg 112, D-21029 Hamburg, Germany}
\altaffiltext{2}{Max-Planck-Institut f\"ur Sonnensystemforschung,
  D-37191 Katlenburg-Lindau, Germany} \altaffiltext{$^\star$}{Marie
  Curie Outgoing International Fellow}




\begin{abstract}
  
  We present the first direct magnetic field measurements on M dwarfs
  cooler than spectral class M4.5. Utilizing a new method based on the
  effects of a field on the Wing-Ford FeH band near 1 micron, we
  obtain information on whether the integrated surface magnetic flux
  ($Bf$) is low (well under 1 kilogauss), intermediate (between 1 and
  about 2.5 kG), or strong (greater than about 3 kG) on a set of stars
  ranging from M2 down to M9. Along with the field, we also measure
  the rotational broadening ($v\,\sin{i}$) and H$\alpha$ emission
  strength for more than 20 stars. Our goal is to advance the
  understanding of how dynamo field production varies with stellar
  parameters for very low-mass stars, how the field and emission
  activity are related, and whether there is a connection between the
  rotation and magnetic flux.
  
  We find that fields are produced throughout the M-dwarfs. Among the
  early M stars we have too few targets to yield conclusive results.
  In the mid-M stars, there is a clear connection between slow
  rotation and weak fields. In the late-M stars, rotation is always
  measureable, and the strongest fields go with the most rapid
  rotators.  Interestingly, these very cool rapid rotators appear to
  have the largest magnetic flux in the whole sample (greater than in
  the classical dMe stars). H$\alpha$ emission is found to be a good
  general proxy for magnetic fields, although the relation between the
  fractional emission and the magnetic flux varies with effective
  temperature.  The known drop-off in this fractional emission near
  the bottom of the main sequence is not accompanied by a drop-off in
  magnetic flux, lending credence to the hypothesis that it is due to
  atmospheric coupling to the field rather than changes in the field
  itself. It is clear that the methodology we have developed can be
  further applied to discover more about the behavior of magnetic
  dynamos and magnetic activity in cool and fully convective objects.

\end{abstract}


\keywords{stars: low-mass, brown dwarfs --- stars: magnetic fields}

\section{Introduction}
\label{sect:Intro}

Magnetic fields are pervasive in astrophysics, and create many
important physical effects, yet they remain one of the more poorly
understood aspects of astrophysical objects. Much of what is
interesting about the Sun on human timescales follows directly from
its production of fields, but even on so nearby an object many
mysteries remain. This is even more true of stars that are not like
the Sun, and the most common type of star (M dwarfs) is the least
understood.

Although magnetic fields play an important role from the beginning
(initially regulating accretion of matter onto a star and regulating
angular momentum and mass loss throughout the star's lifetime), they
are very hard to measure directly. We rely instead primarily on proxy
measures due to the heating of stellar atmospheres caused indirectly
by the magnetic fields. These include the strength of various emission
lines, the total X-ray flux, and other radiative markers. To directly
measure fields, the Zeeman effect on spectral lines has been utilized
\citep{Saar01}. For spectral types A to mid-M, this can be done with
atomic lines. Sometimes field maps can be produced through Zeeman
Doppler imaging \citep[see][ for the first example of this on an M
dwarf]{Donati06}. The few low-mass cases analyzed have shown that the
proxies appear to be good substitutes for field measurements
themselves. That has not been established for the very low-mass stars
(VLMS) -- objects of spectral class M5 or later -- as there have been
no direct field measurements on them.  The primary result to date of
direct field measurements (utilizing atomic lines) on earlier M dwarfs
has been the detection of strong fields on dMe stars
\citep[eg.][]{JKV96}.

For the cooler half of M dwarfs (and any cooler objects), atomic lines
lose their utility, since the only optical/IR lines left must be very
strong to be visible against the haze of molecular features.  This
means they are dominated by pressure broadening, and Zeeman broadening
is masked. Molecules themselves can be sensitive to magnetic fields,
although the theoretical understanding of the spectral effects is much
less developed. The behavior of the Wing-Ford band of FeH (near one
micron) in sunspot spectra suggests that it displays useful
information on magnetic fields.  \citet[][hereafter
Paper~I]{Reiners06a} developed an observational technique for
extracting field information from this diagnostic. In this paper, we
utilize that method to provide the first substantial sample of direct
field measurements for VLMS.

The VLMS pose a number of unique questions for the overall
understanding of stellar activity. They are unquestionably fully
convective, so they clearly cannot support an interface $\alpha
\omega$-dynamo as is thought to operate in the solar case. It is known
that they have longer spindown times \citep[e.g., ][]{Mohanty03,
  Reiners06b}, which implies less efficient magnetic braking. This
could be due either to weaker fields, weaker magnetic winds (due to
the field configuration or less acceleration due to weaker coronal
heating), the fully convective nature of the objects (if radiative
cores are not spun down in more massive stars), or some combination of
these.

The convective overturn times grow longer for lower luminosity objects
while their rotation periods tend to get shorter \citep[e.g.,
][]{Mohanty03, Bailer04}, so the Rossby number (which is the period
divided by the overturn time) can become rather small. This number is
relevant to the activity levels in solar-type stars \citep{Noyes84}.
It will be relevant in convective dynamos if the scale of dynamo
action is large enough. The fact that the Rossby number is getting
rather small makes this scale smaller, which could allow rotation to
still play a role (as it would in the mean-field case of an $\alpha
^2$ dyamo). If the dynamo scale gets too small (as in the so-called
fully turbulent dynamo), then rotation should not matter as much
\citep{Durney93, Dobler06}. It is also unclear how much small-scale
fields will cascade to larger scales, and how global or local the
surface field distribution will become. This geometry, in turn, can
affect the efficiency of magnetic braking for the VLMS.

What is known is that for the coolest M dwarfs, very rapid rotation is
not generally accompanied by very high activity \citep{BM95, Mohanty03}.  
A dropoff in activity is found in late M stars \citep{Gizis00}, 
yielding few stars with H$\alpha$ emission cooler
than spectral type M9. One explanation for this is the extreme
neutrality of the photospheres \citep{MBS02}. This prevents turbulence
from twisting the field into non-potential configurations (the basic
mechanism for stellar magnetic activity). An alternative explanation
is that the field itself is disappearing. The appearance of radio (and
sometimes optical or X-ray) flares on some ultracool objects argues
for the continued presence of a magnetic field
\citep[eg.][]{Berger06}, but direct measurement of magnetic fields in
VLMS is still missing.  Direct measurement also allows a quantitative
comparison of the fields on objects with different rotations and
luminosities, which should yield insights into the dynamo processes.

\section{Data}
\label{sect:Data}

We begin our foray into this largely unexplored territory by studying
some of the brighter and more slowly-rotating examples of mid- and
late-M dwarfs.  These targets are easiest for our method of finding
fields yield to good results on, so this observational sample is
definitely biased, and cannot be used to draw conclusions about the
full range of M dwarfs. In particular, we have avoided the really
rapid rotators among the late-type stars for now, so the behavior of
fields at high rotation is not tested here. Our 10 stars that are
earlier than M5 are in territory that has been explored previously
with atomic Zeeman broadening; they serve to test whether our method
yields results consistent with earlier work. Some of them show
H$\alpha$ emission, some don't.  It is actually hard to find nearby
VLMS without H$\alpha$ emission between M5 and M9 \citep{West04}, and
all but 2 of our 14 targets in that range exhibit it (but at different
levels). The ubiquity of this emission is partly due to the ease of
seeing plasma at chromospheric temperatures against the very cool
photospheres, but of course without a chromosphere (which implies
non-radiative heating) there would be no H$\alpha$ at all in these
very cool objects. The list of targets appears in
Table\,\ref{tab:observations}.

Our data were taken with HIRES at Keck~I during three observing runs
on March~1, August~14 and 15, and December~18, 2005. We chose a
similar setup in all three nights that covers the wavelength region
from 5600 to 10\,000\,\AA; order coverage is incomplete in the red.
Our setup was chosen to simultaneously obtain spectra in the
wavelength region around H$\alpha$ and at 1 micron, where a strong FeH
band appears in spectra of very cool stars.  To minimize light loss
and to reach a reasonable signal-to-noise ratio (SNR) even in our
faintest targets, we chose a slit width of 1.15\,arcsec yielding a
resolution of $R \approx 31\,000$. This resolution is enough to detect
signatures of magnetic fields in the FeH lines, as was shown in
Paper~I. The targets we observed and exposure times we used are given
in Table\,\ref{tab:observations}.

Data were reduced in a standard fashion using MIDAS routines. They
were flat-fielded and filtered for cosmic rays. We subtracted light
from sky emission lines by individually extracting the sky spectrum. 
Our slit length of 14\arcsec\ provided enough room on the chip for the
reduction of a sky spectrum on both sides of the target's spectrum. 
We interpolate this across the target to provide sky subtraction. We
observed several telluric standard stars during all nights. At the
region of the FeH lines, telluric absorption features prove not to be
important in the spectra taken at the high altitude of Keck
Observatory, so we did not subtract any telluric features.

\placetable{tab:observations}

In the spectral region of the FeH band our targets are much brighter
than in the bluer optical wavelength range. Although detector
efficiency goes down at such long wavelengths, the overall signal is
still very high compared to the bluer part of the spectrum. Thanks to
the new thick high-resistivity CCD the system throughput is of the
order of 2\,\% at 1\,$\mu$m (before the detector upgrade it was
approximately 1\,\%).  We measured the SNR of our data in the FeH band
mainly in two short sections where flux reaches the continuum in
almost all spectra. The two sections are $\lambda = [9931.2 -
9932.0]$\,\AA\ and $\lambda = [9937.0 - 9938.1]$\,\AA. The SNR we
measure for our data is given in column 5 in
Table\,\ref{tab:observations}; it is in the range between 20 and 150.

\section{Analysis}

For our analysis of magnetic activity in ultracool objects, we measure
emission in the H$\alpha$ line, the projected rotational velocity
$v\,\sin{i}$, and the strength of the magnetic flux $Bf$ as explained
in the following.

\subsection{H$\alpha$ activity}

Emission in the H$\alpha$ line is thought to be a good proxy for
stellar magnetic activity. To measure the equivalent width in the
H$\alpha$ line against the continuum, we normalize the line at two
footpoints blue- and redwards of H$\alpha$. The footpoints are the
median values at 6545 -- 6559\,\AA\ on the left hand side, and at 6567
-- 6580\,\AA\ on the right hand side of the H$\alpha$ line.  None of
the emission lines found in our targets extends into the region used
for normalization. The H$\alpha$ equivalent width is then measured by
integrating the flux from 6552 to 6572\,\AA. We convert the measured
H$\alpha$ equivalent width into H$\alpha$ flux at the star by
measuring the flux per unit equivalent width from the continuum flux
in synthetic PHOENIX spectra \citep[][we used the DUSTY models at low
temperatures]{Allard01}. The model temperature was taken from the
spectral types of our targets and calculated according to the
conversions given in \cite{Kenyon95} for objects earlier than spectral
type M7, and in \cite{Golimowski04} in later stars.  The same
temperature was used to derive the bolometric flux $F_{\rm bol} =
\sigma T^4$ and to calculate the ratio of H$\alpha$ luminosity and
bolometric luminosity from $L_{\rm H_{\alpha}}/L_{\rm bol} = F_{\rm
  H_{\alpha}}/F_{\rm bol}$.

\subsection{Rotation velocity}

The absorption lines we use for our analysis are all due to the same
rovibrational band of FeH. We found no indication for CrH in our
targets (Paper~I). Thus, we expect the structure of the absorption
band to be the same in all targets except for its strength. As we have
shown in Paper~I, we found that the strength \emph{a} of the FeH
absorption band can be accurately described by an optical depth
scaling. This strategy accounts for saturation of the strong FeH
absorption.

For most of our target M-stars, rotation velocities have been
determined previously by \cite{Delfosse98}, \cite{Mohanty03} and
\cite{Bailer04}.  These authors measure the broadening of the spectra
in comparison to a template spectrum via the cross-correlation
technique. For the comparison template, the latter two use the slow
rotator Gl\,406 (M5.5); \cite{Bailer04} also calculates rotation
velocities from a comparison to 2MASS\,1439 (L0.0) although
$v\,\sin{i} \approx 10$\,km\,s$^{-1}$ is reported for it. Both papers
point out that the use of a template spectrum of different spectral
type may introduce a systematic error in the determination of
$v\,\sin{i}$, which is especially important in the latest M-dwarfs.
Their spectra are significantly different when compared to the mid
M-type spectrum of Gl\,406.

The FeH band around 1~micron is an ideal region to determine rotation
velocities in the coolest M-type dwarfs. It contains a number of
narrow spectroscopic lines. Templates of very cool objects can be
constructed from warmer stars, it is even possible to reproduce the
spectrum of a moderately rotating L-type dwarf by artificially
enhancing the strength of the FeH absorption lines in the spectrum of
a slowly rotating mid-M dwarf (Paper~I). To determine the rotation
velocity, we first construct a template spectrum of a slow rotator
that has the same FeH band strength as the target spectrum. After that
we artificially broaden the constructed template spectrum in order to
fit the FeH band of our target spectra. With this strategy we minimize
uncertainties that arise from differences between the template and the
target spectrum.

For each of our targets, we search for the values of FeH strength $a$
and projected rotation velocity $v\,\sin{i}$ that provide the best
$\chi^2$ fit to our spectrum. The full wavelength region 9893 --
9997\AA\ is used for a robust estimate of $v\,\sin{i}$ and $a$ (for our
final value of $v\,\sin{i}$ we include the magnetic flux $Bf$ as third
parameter as described in the next chapter). As a template, we use the
spectrum of GJ~1002 \citep[M5.5, no H$\alpha$ emission, $v\,\sin{i} <
2.3$\,km\,s$^{-1}$;][]{Delfosse98}.  The behavior of FeH absorption
strength with spectral type is consistent with results from low
resolution spectroscopy in larger samples (Paper~I). In addition to an
accurate determination of the projected rotation velocity, this
strategy enables us to directly compare an artificial template
spectrum of a magnetically inactive non-rotating ``star'' to the
target spectra. Thus we can determine line broadening from direct
comparison as is sometimes done in hotter stars where individual
atomic lines are available in the spectra.

At the resolution of our data ($R \approx 31\,000$; FWHM~$\approx
9.5$\,km\,s$^{-1}$) we estimate that we can detect rotation velocities
of about 3\,km\,s$^{-1}$ or more. We tested the sensitivity of our
method to small values of $v\,\sin{i}$ by comparing $\chi^2$ of our
best fit to the spectrum of Gl\,406 ($v\,\sin{i} = 3$\,km\,s$^{-1}$)
to the best model with zero rotation; the value of $\chi^2$ is
significantly larger (more than 2\,$\sigma$) in the model with zero
rotation. However, our method is severly limited by systematic
uncertainties (most significantly the rotation velocity of the
template stars), and we consider a detection of $v\,\sin{i} =
3$\,km\,s$^{-1}$ marginally significant. Since the average seeing at
Keck observatory is around $~0.8$\arcsec\ (FWHM~$\approx
6.5$\,km\,s$^{-1}$), we checked the stability of our resolution using
the telluric oxygen A-band. We found that in all our data the
resolution is constant; variations of the resolution between different
data do not exceed 5\%.  Thus, we did not apply any correction before
determining rotational broadening.  Since the rotational velocity of
GJ\,1002 is only known as an upper limit of 2.3\,km\,s$^{-1}$, and due
to the limited resolution, we cannot accurately determine rotation
velocities in the slowest rotators.  The uncertainty of the results
also depends on the saturation of the absorption bands, since blending
becomes an issue when the FeH lines become stronger and broader. This
is more important in the cooler targets where we also see only rapid
rotators. We estimate the uncertainty of our $v\,\sin{i}$ measurements
to be on the order of 20\,\% in all our objects.

\subsection{Magnetic field}

In Paper~I we introduced a method to determine the surface magnetic
flux in VLMS from the magnetically sensitive FeH lines at around 1
micron. Land\'e-$g$ factors for FeH are not known and it is not yet
possible to calculate the magnetic splitting of FeH lines.  However,
magnetic splitting can be observationally characterized by comparison
to template spectra of stars with known magnetic fields.  In Paper~I,
we investigated the spectra of a magnetically inactive star and a
magnetically active star, and we identified lines that are
particularly insensitive or sensitive to magnetic splitting.  In order
to estimate the total magnetic flux, we use a linear interpolation
between two reference spectra. In this work, we use the spectra of
Gl\,873 ($S_{\rm Gl\,873}$) and GJ\,1002 ($S_{\rm GJ\,1002}$) as
templates. The former shows a magnetic field strength of $Bf \approx
3.9$\,kG \citep{JKV96} measured from atomic lines (our spectrum also
covers the line used in that work, and we see no significant signs of
variation). GJ\,1002 shows no activity and the magnetically sensitive
lines are among the narrowest in our sample targets; we assume a zero
field in GJ\,1002.  Under the assumption that the magnetic field $B$
is similar in the magnetic reference star and the target star, the
interpolation factor $p$ is linear in $f$ (or $Bf$).  Thus, the value
of $Bf$ can be estimated from interpolating $S_{\rm new}$ according to
\begin{equation}
\label{eq:S}
S_{\rm new} = pS_{\rm Gl\,873} + (1-p) S_{\rm GJ\,1002}.
\end{equation}
The estimate of $Bf$ in this interpolation is then $p \cdot Bf({\rm
  Gl\,873}) = p \cdot 3.9$\,kG. \footnote{We chose GJ\,1002 as a
  reference instead of GJ\,1227 (which we used in Paper~I) since the
  FeH lines in the former are stronger. The spectra of GJ\,1227 and 
  GJ\,1002 show no significant difference except for their strength.}

We use the same normalization of FeH band strength $a$ as in Paper~I,
i.e. unity depth of the FeH absorption is defined by the spectrum of
GJ\,1227. To calculate an interpolation, first the FeH lines are
enhanced in strength. The two enhanced templates are then co-added
according to Eq.\,\ref{eq:S} and rotational broadening is applied. The
magnetic field values $B$ of Gl\,873 and the target can not be
expected to be similar, but the strong broadening of magnetically
sensitive FeH lines allows a very robust estimate of the integrated
flux $Bf$ as was demonstrated in Paper~I. We conservatively estimate
the accuracy of such a detection to be on the level of a kilogauss.
The main source of uncertainty is the lack of knowledge about
splitting patterns in the molecular FeH lines, which makes it
necessary to assume similar magnetic fields $B$. It is not possible to
disentagle the parameters $B$ and $f$.  Nevertheless, our empirical
approach enables us to reliably distinguish between the cases of a
very weak or no field, a mean field comparable to the one on Gl\,873
($\sim 4$\,kG), or an intermediate mean field closer to $Bf = 2$\,kG.

\placetable{tab:mag_regions}

In order to determine the magnetic flux $Bf$ from magnetically
sensitive lines, we focus on four spectral regions that are
particularly useful for this purpose.  We use different regions for
slow rotators and for rapid rotators, they are listed in
Table\,\ref{tab:mag_regions}. We search for the interpolation between
the two reference spectra that best matches our target spectrum by
simultaneously fitting for the parameters $a$, $v\,\sin{i}$ and $Bf$
in these four regions using $\chi^2$-minimization. Our final values of
$a$ and $v\,\sin{i}$ are the ones from fits in the four regions
including the effect of magnetic broadening. The best fit field found
depends on the values of $a$ and $v\,\sin{i}$, however, forcing a
variation in the magnetic flux does not significantly affect the other
two parameters without worsening the fit.

In the majority of our spectra, direct $\chi^2$-fitting provides
reliable results in the sense that the interpolation matches the
target spectrum over the whole region investigated. In particular, it
fits both, magnetically sensitive \emph{and} insensitive lines. In two
cases, the difference between the interpolated spectrum and the target
spectrum cannot be explained by noise and $\chi^2$ fitting does not
provide reliable results. For those we tried to estimate the 
magnetic flux from ratios between magnetically sensitive and
insensitive lines as explained in Paper~I.  We will individually
discuss these two in the following section.

We consider a measurement of $Bf$ significant in stars where the
interpolation provided a good fit with $\chi_{\rm min}^2 \approx 1$,
i.e. where the difference can be explained with photon noise. For
these cases, we estimate the possible range of magnetic fluxes ($Bf
\pm \Delta Bf$) by searching for the lowest and highest values of $p$
for which $\chi^2 < \chi_{\rm min}^2 + 4$ while varying all other
parameters \citep[i.e., uncertainties on a 2$\sigma$-level, ][
Ch.15]{NR}. These formally derived uncertainties are given in
Table\,\ref{tab:Results}. However, we emphasize that the uncertainties
in our determination of the mean magnetic field are dominated by
systematic uncertainties as discussed above, and we estimate the
accuracy of our mean magnetic field measurements to be on the order of
a kilogauss.

\section{Results}
\label{sect:results}

\begin{figure}
\vspace*{-25mm}
  \includegraphics[width=\hsize,clip=,bbllx=0,bblly=190,bburx=648,bbury=468]{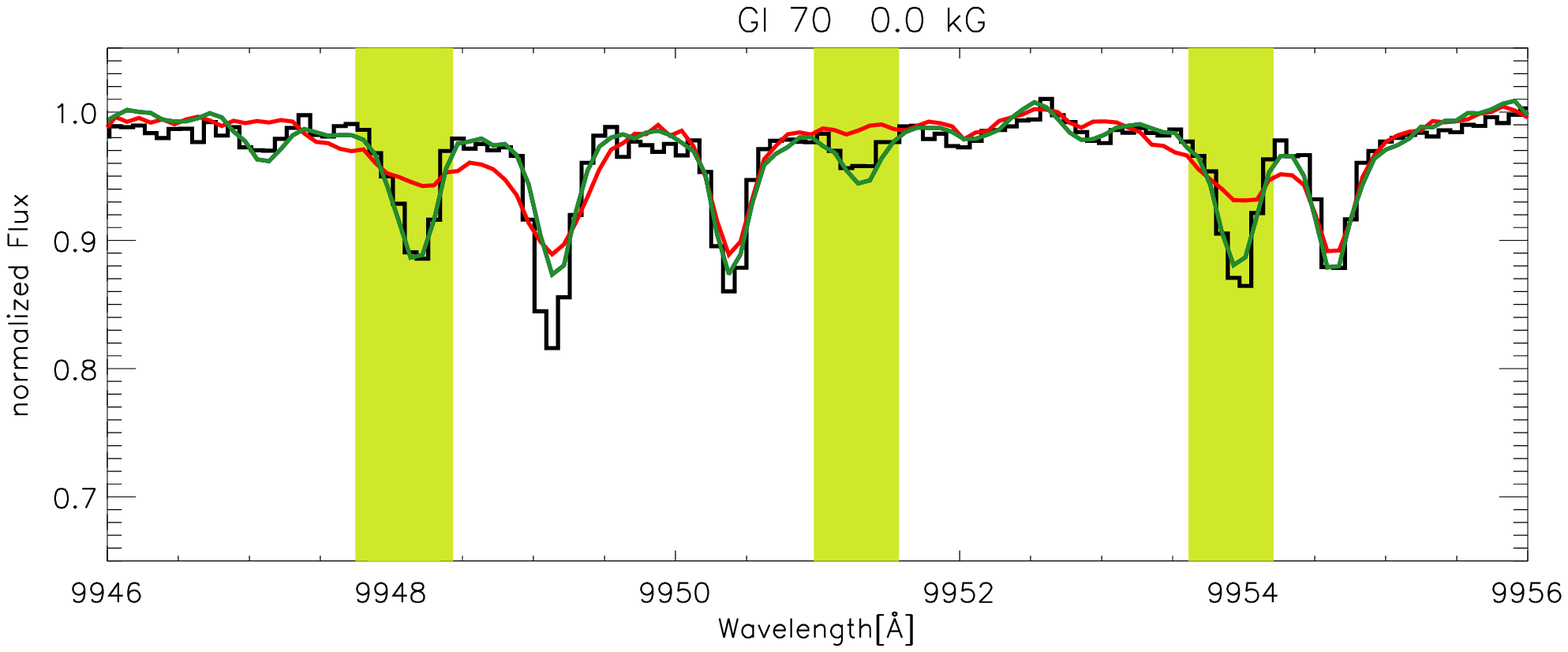}
  \includegraphics[width=\hsize,clip=,bbllx=0,bblly=190,bburx=648,bbury=468]{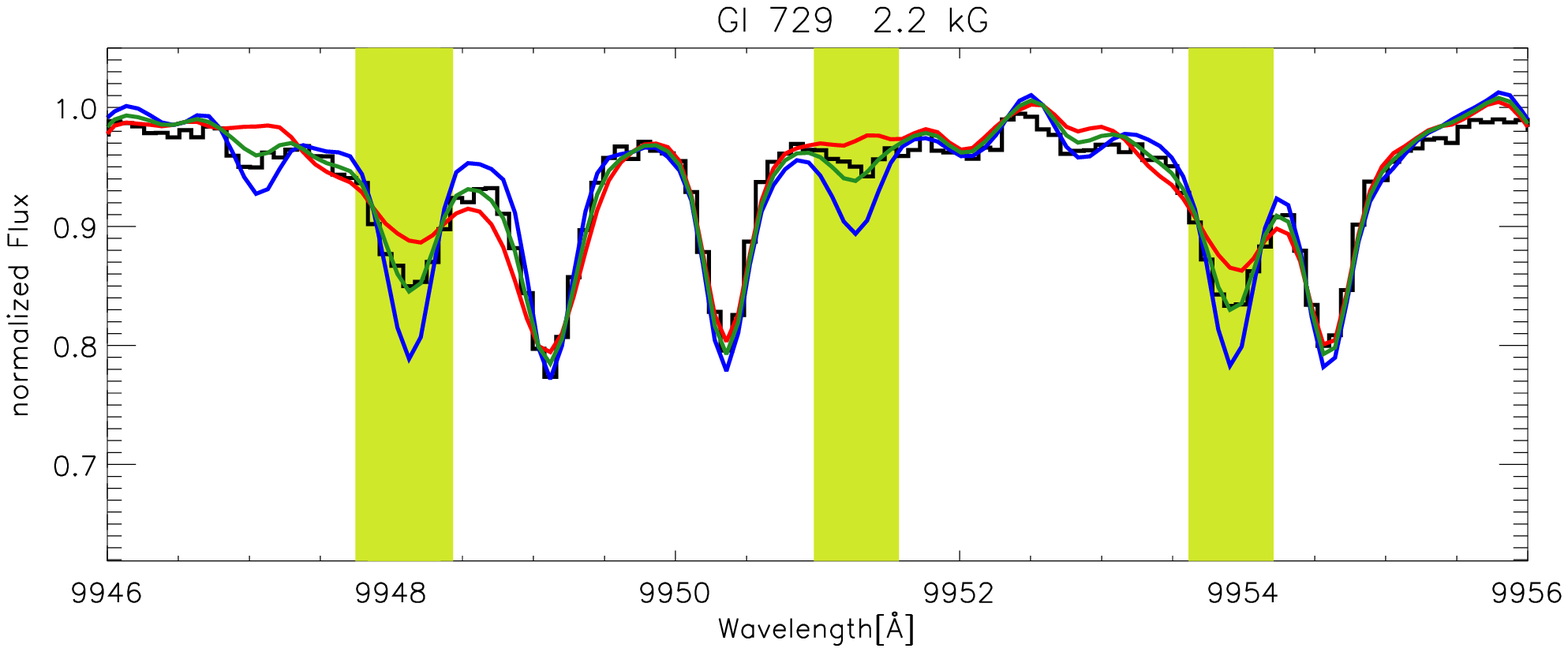}
  \includegraphics[width=\hsize,clip=,bbllx=0,bblly=190,bburx=648,bbury=468]{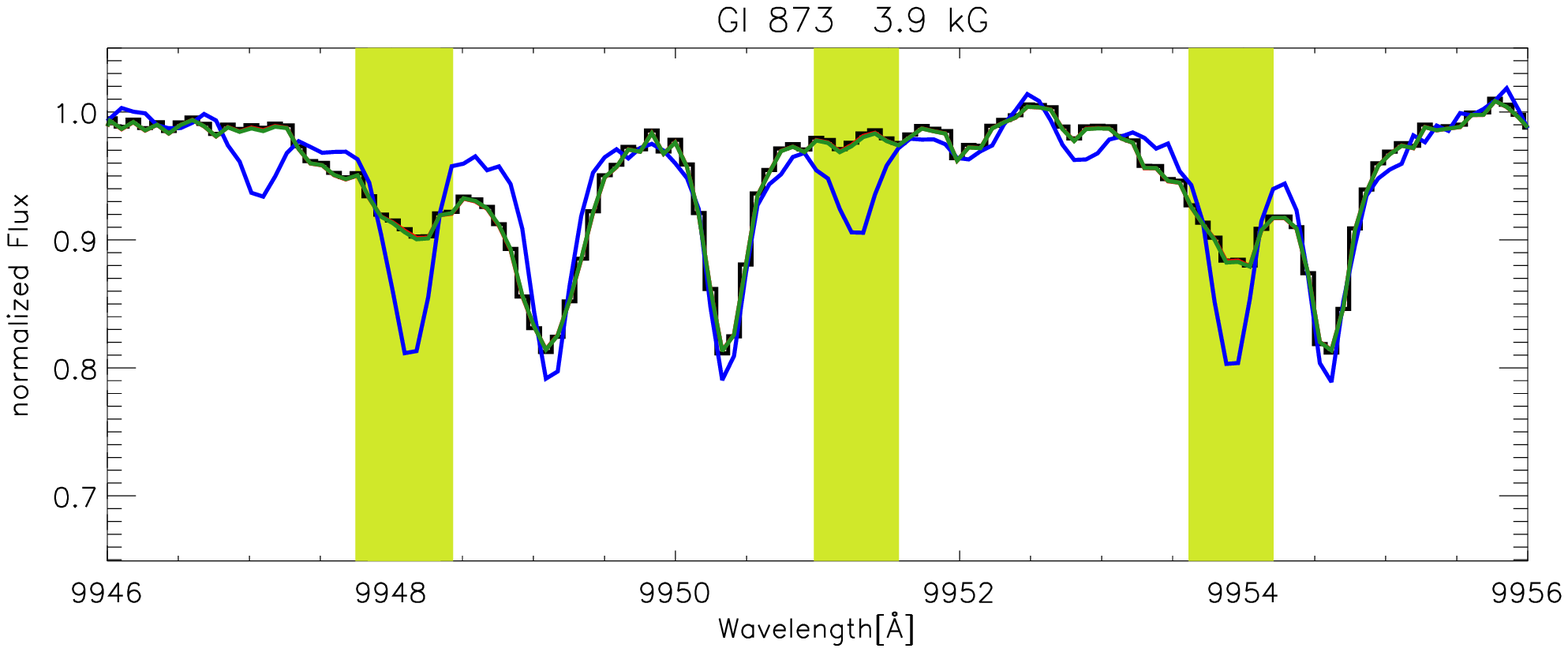}
  \caption{\label{fig:fit1}A section of the fits to Gl~70 (M2), Gl~729 (M3.5), and Gl~873 (M3.5). In all plots the case of no magnetic flux is shown in blue, strong magnetic flux ($\sim$4\,kG) in red, and the best fit is overplotted in green with the actual value of $Bf$ given above the plots. The green bars indicate parts of the spectrum that are particularly sensitive to the presence of a magnetic field. There is no red line for Gl 873 because the green line
is coincident with it.}
\end{figure}

\begin{figure}
  \includegraphics[width=\hsize,clip=,bbllx=0,bblly=190,bburx=648,bbury=468]{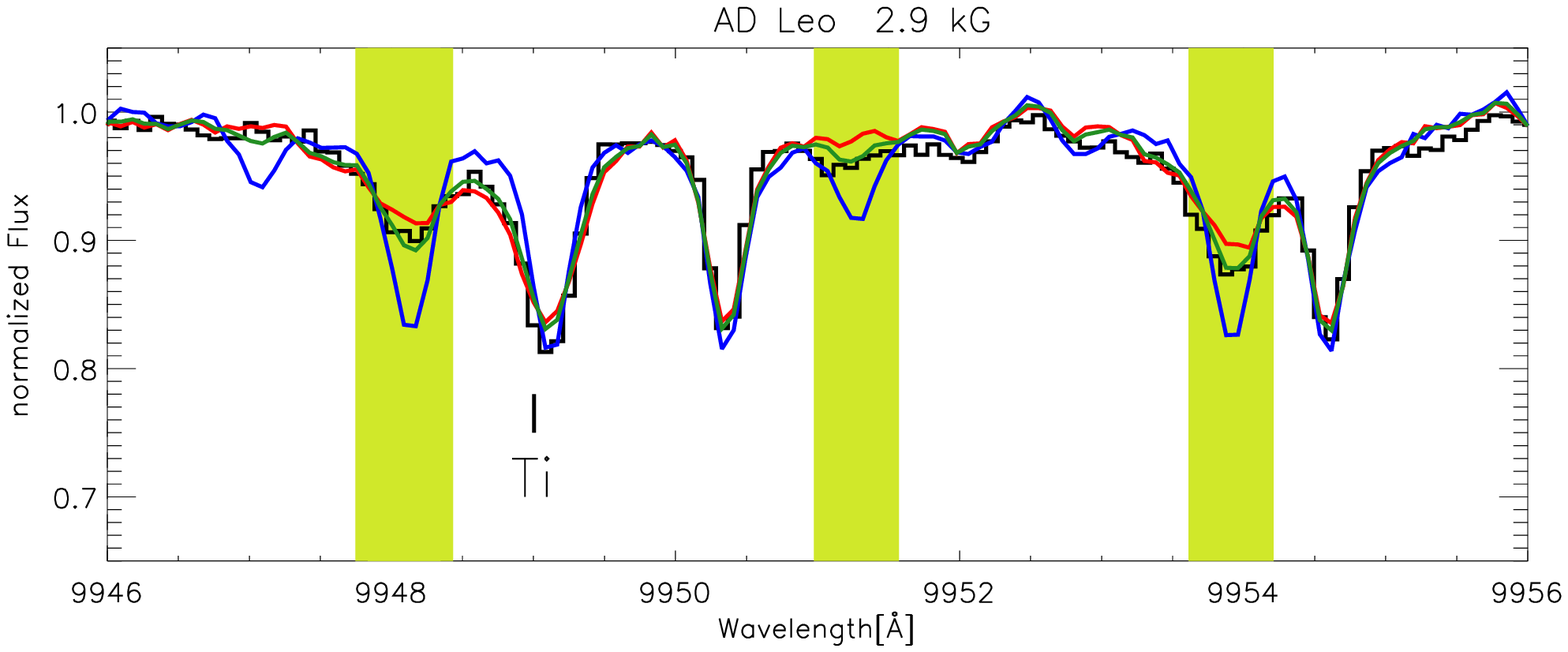}
  \includegraphics[width=\hsize,clip=,bbllx=0,bblly=190,bburx=648,bbury=468]{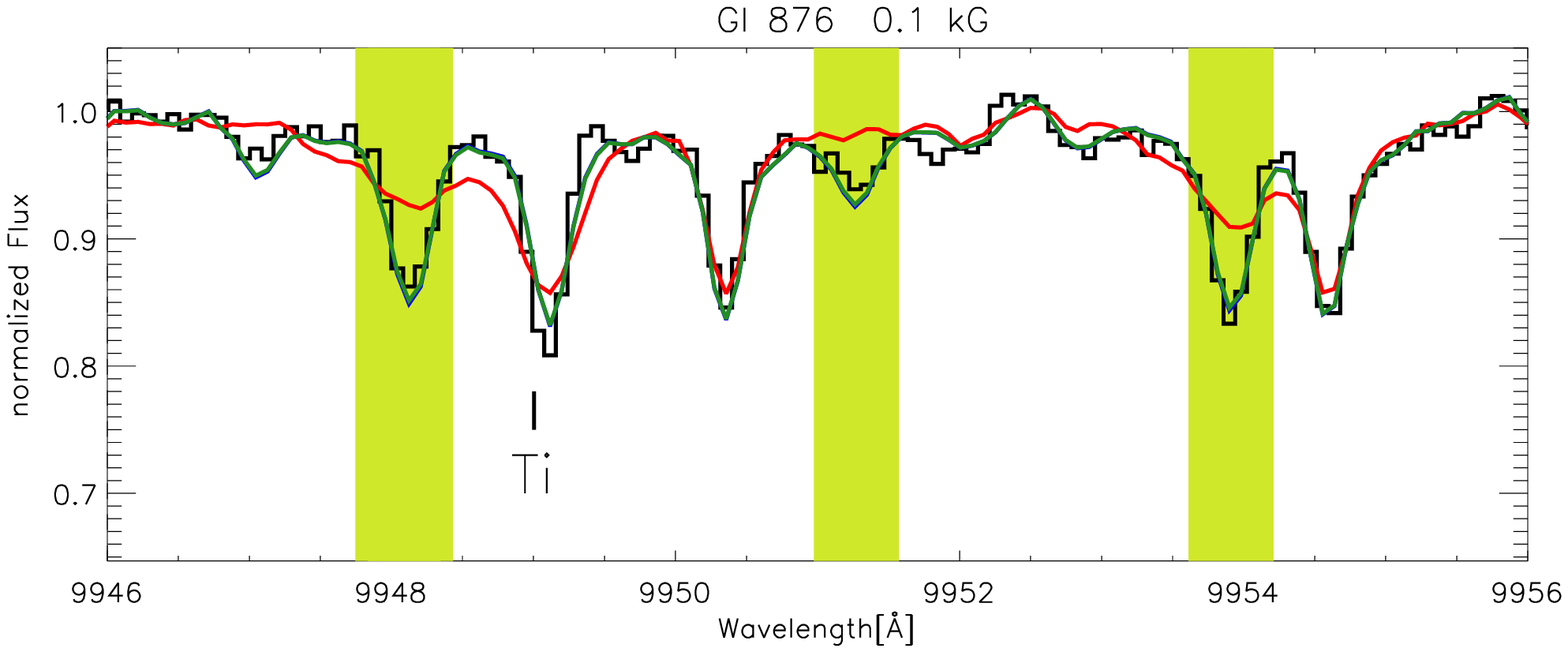}
  \includegraphics[width=\hsize,clip=,bbllx=0,bblly=190,bburx=648,bbury=468]{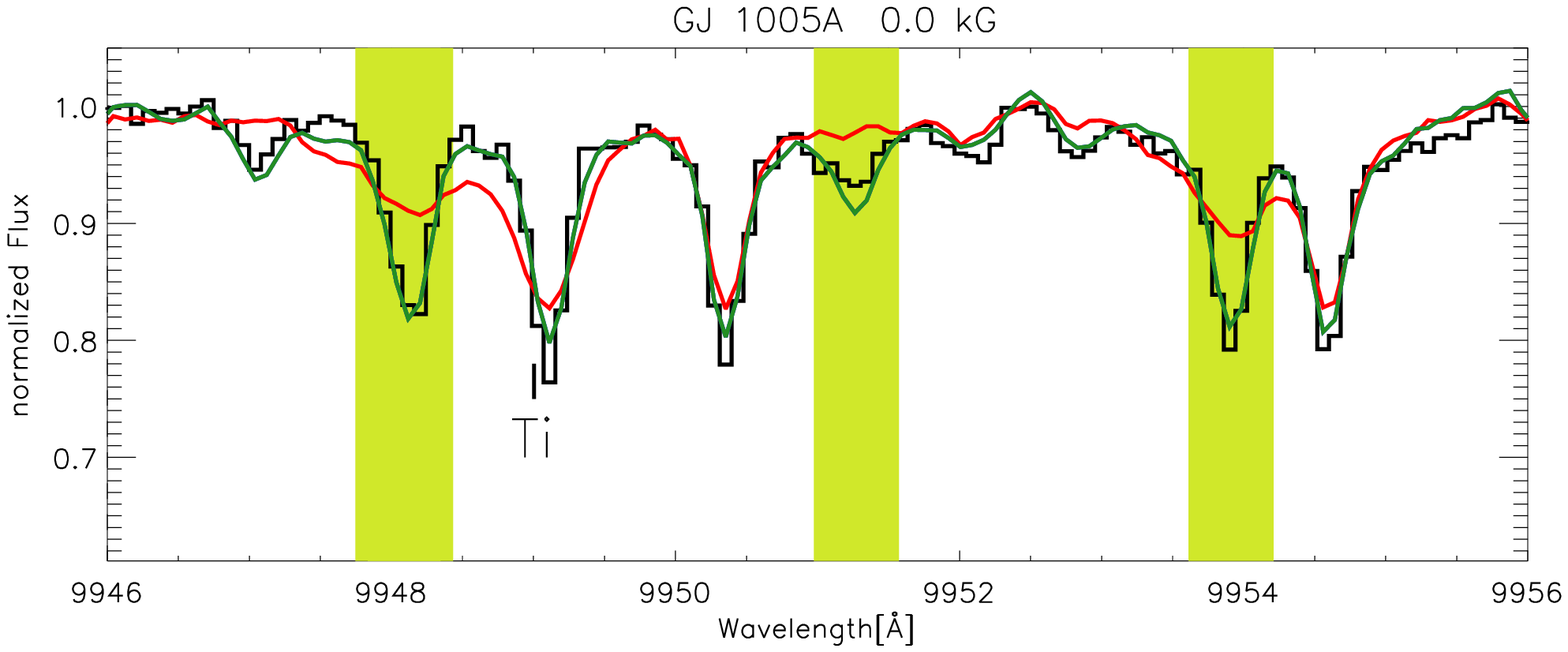}
  \caption{\label{fig:fit2}Same as Fig.\,\ref{fig:fit1} for AD~Leo (M3.5), Gl~876 (M4), and GJ~1005A (M4).}
\end{figure}

\begin{figure}
  \includegraphics[width=\hsize,clip=,bbllx=0,bblly=190,bburx=648,bbury=468]{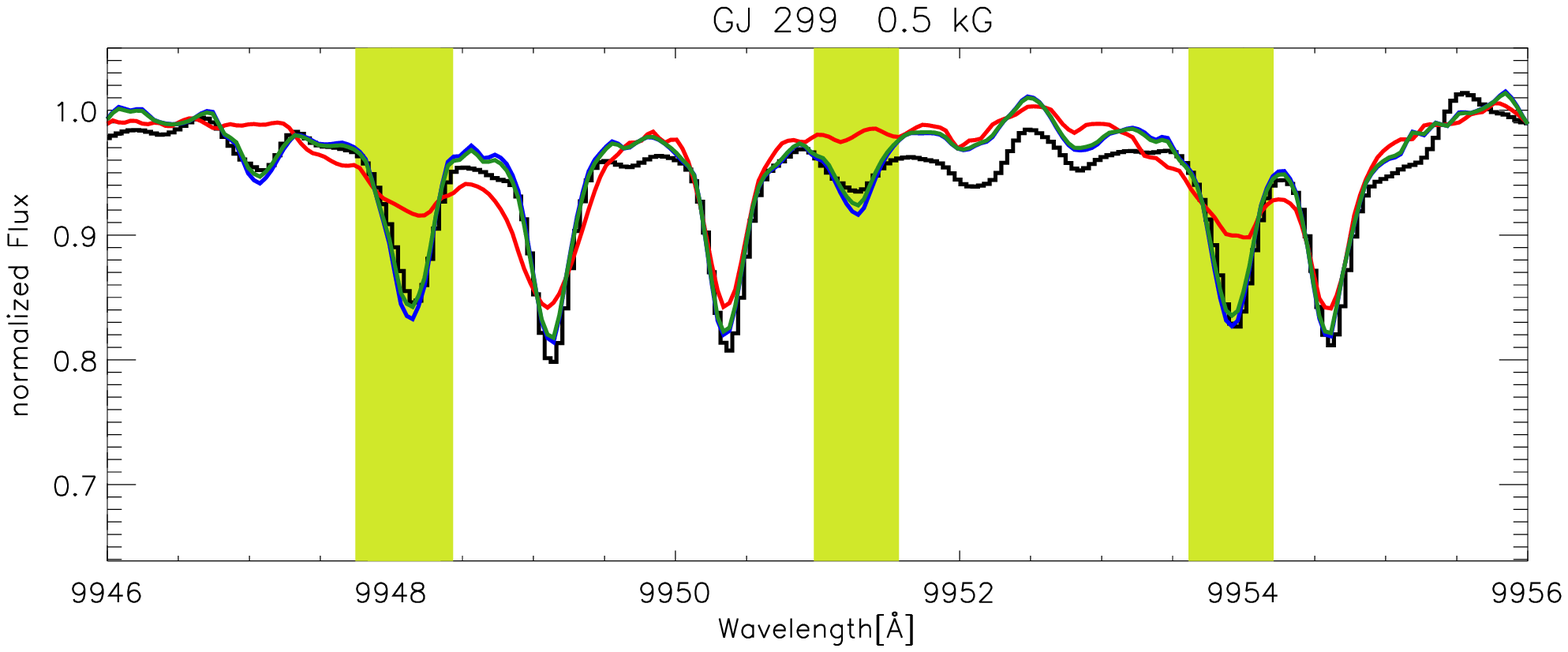}
  \includegraphics[width=\hsize,clip=,bbllx=0,bblly=190,bburx=648,bbury=468]{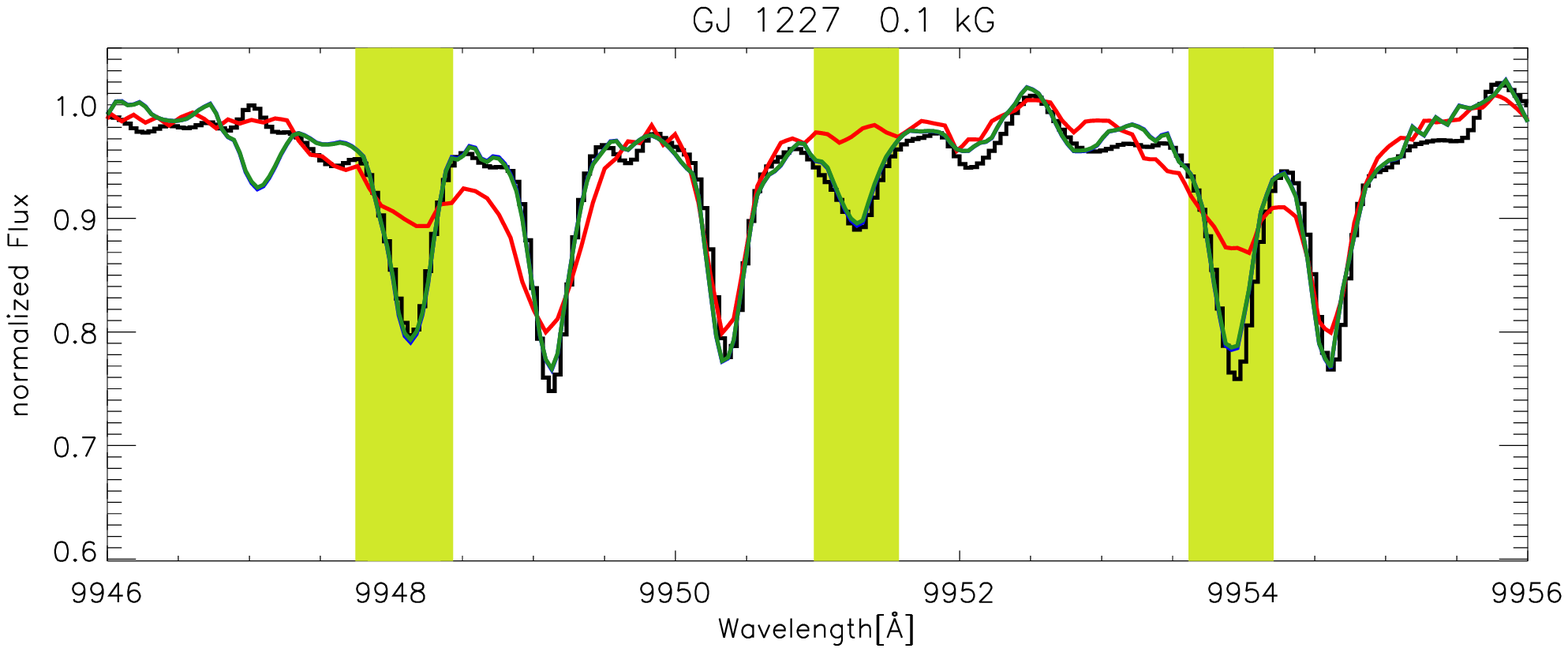}
  \includegraphics[width=\hsize,clip=,bbllx=0,bblly=190,bburx=648,bbury=468]{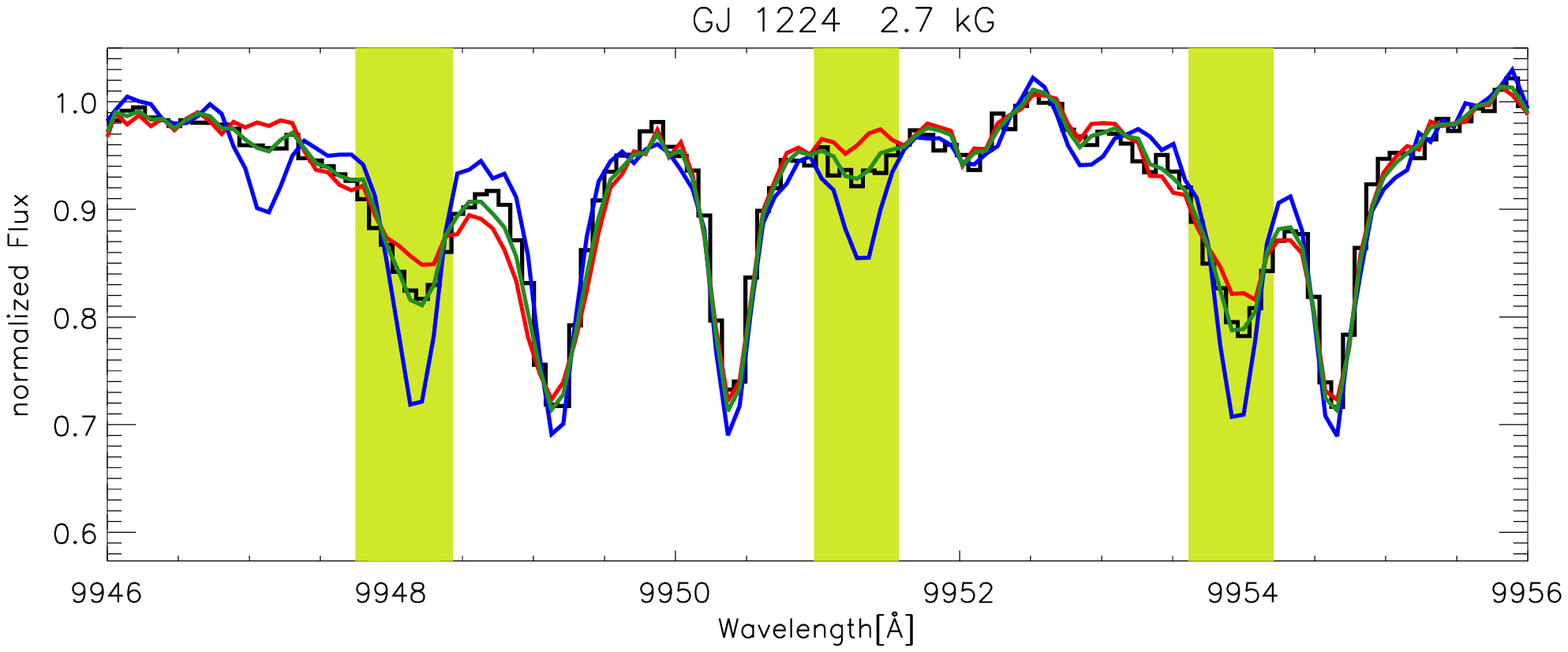}
  \caption{\label{fig:fit3}Same as Fig.\,\ref{fig:fit1} for Gl~299 (M4.5), GJ~1227 (M4.5), and GJ~1224 (M4.5).}
\end{figure}

\begin{figure}
  \includegraphics[width=\hsize,clip=,bbllx=0,bblly=190,bburx=648,bbury=468]{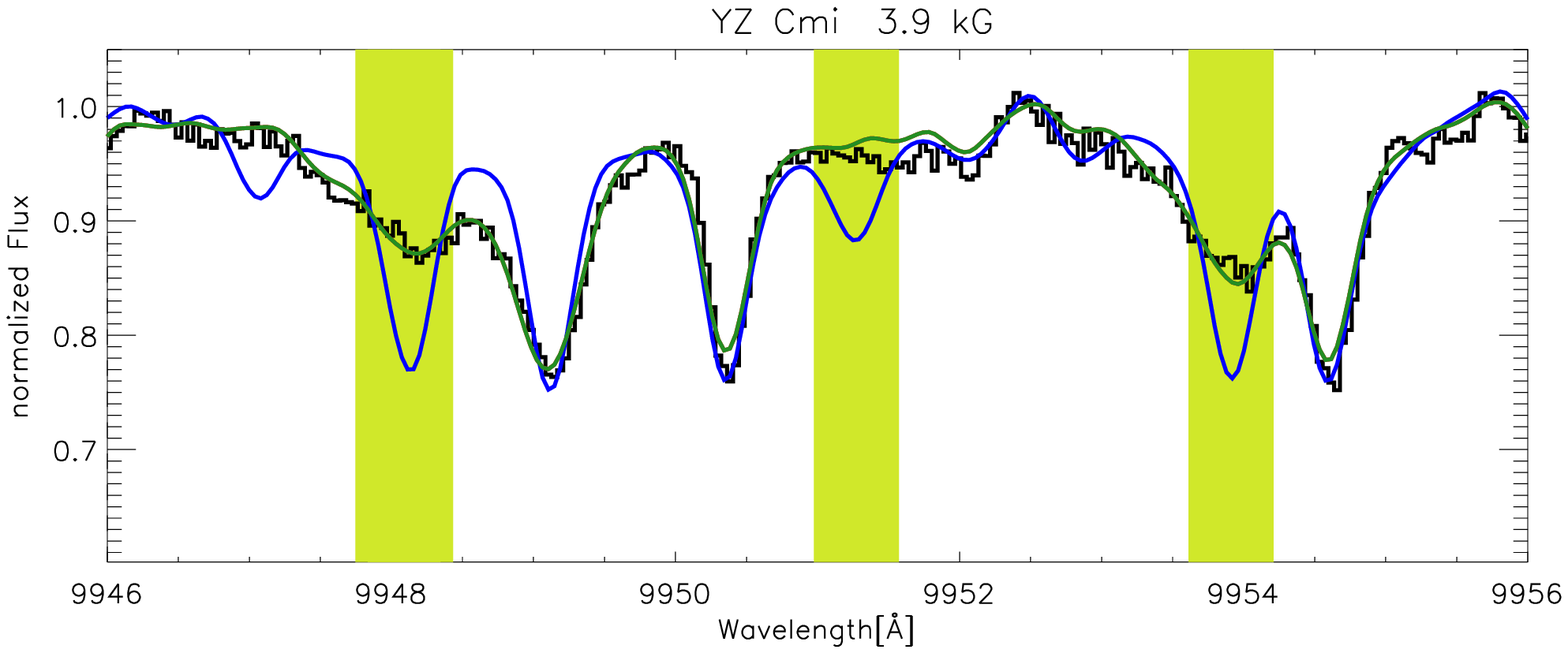}
  \includegraphics[width=\hsize,clip=,bbllx=0,bblly=190,bburx=648,bbury=468]{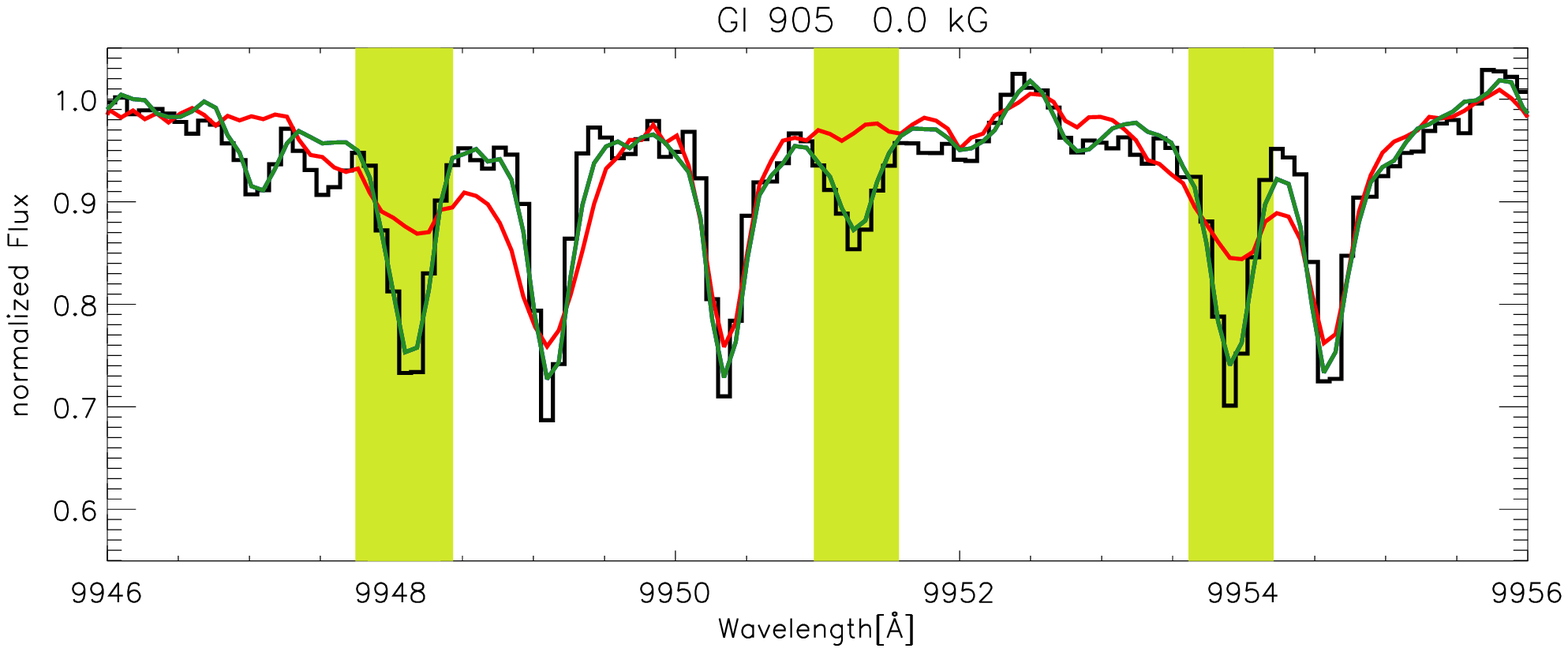}
  \includegraphics[width=\hsize,clip=,bbllx=0,bblly=190,bburx=648,bbury=468]{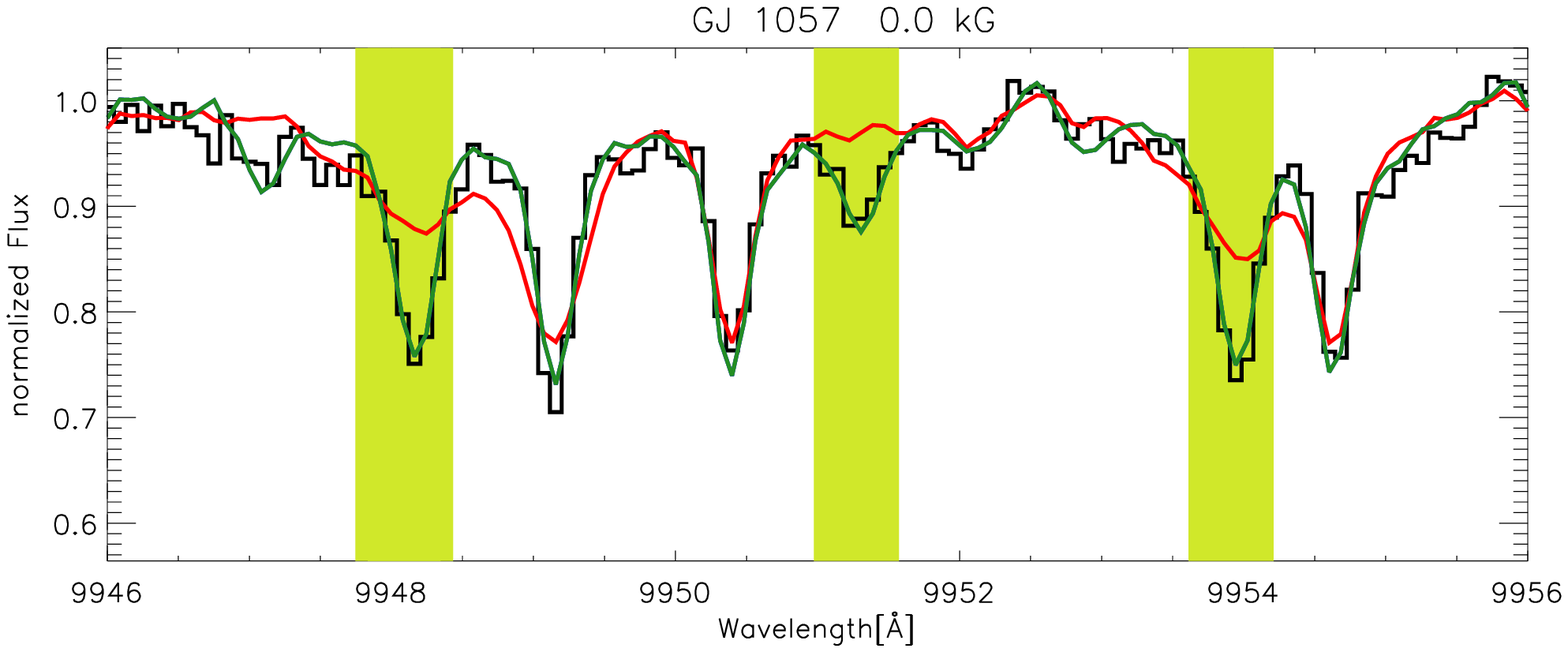}
  \caption{\label{fig:fit4}Same as Fig.\,\ref{fig:fit1} for YZ~Cmi (M4.5), Gl~905 (M5), and GJ~1057 (M5).}
\end{figure}

\begin{figure}
  \includegraphics[width=\hsize,clip=,bbllx=0,bblly=190,bburx=648,bbury=468]{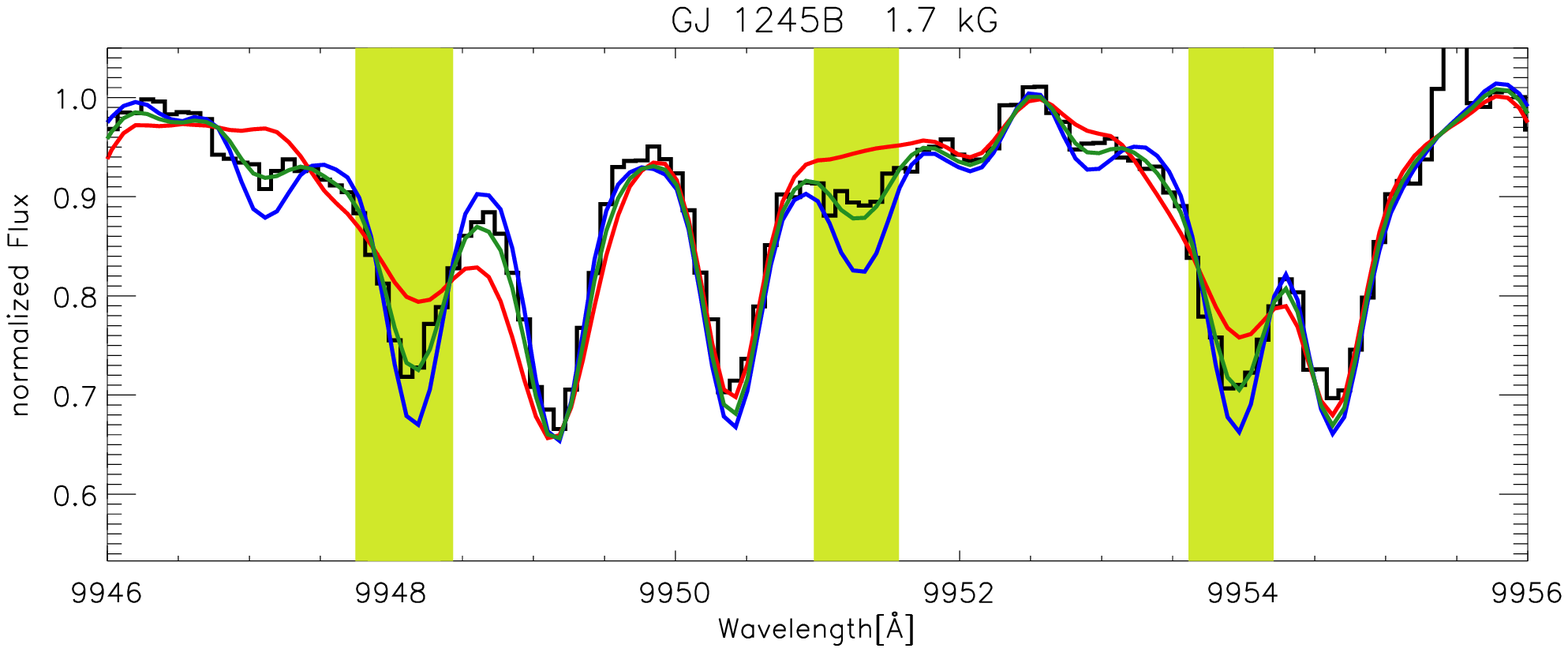}
  \includegraphics[width=\hsize,clip=,bbllx=0,bblly=190,bburx=648,bbury=468]{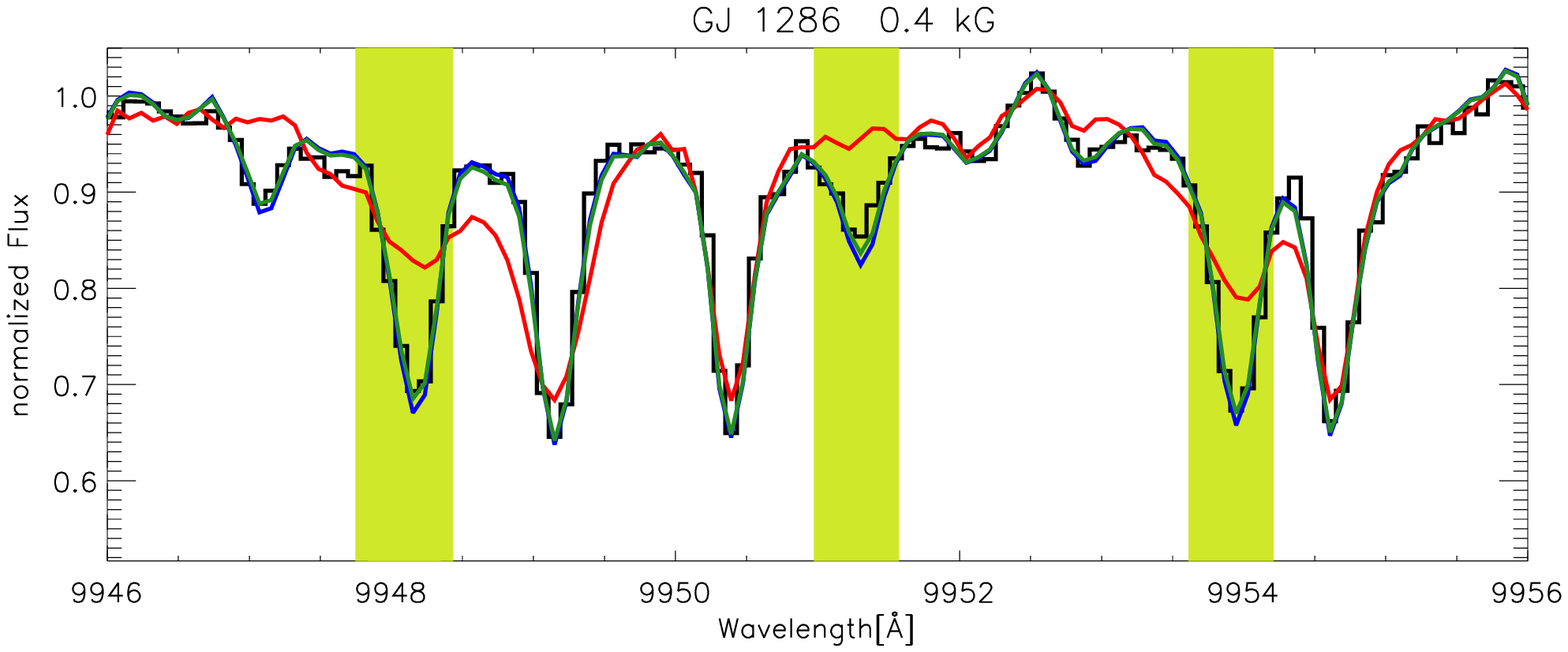}
  \includegraphics[width=\hsize,clip=,bbllx=0,bblly=190,bburx=648,bbury=468]{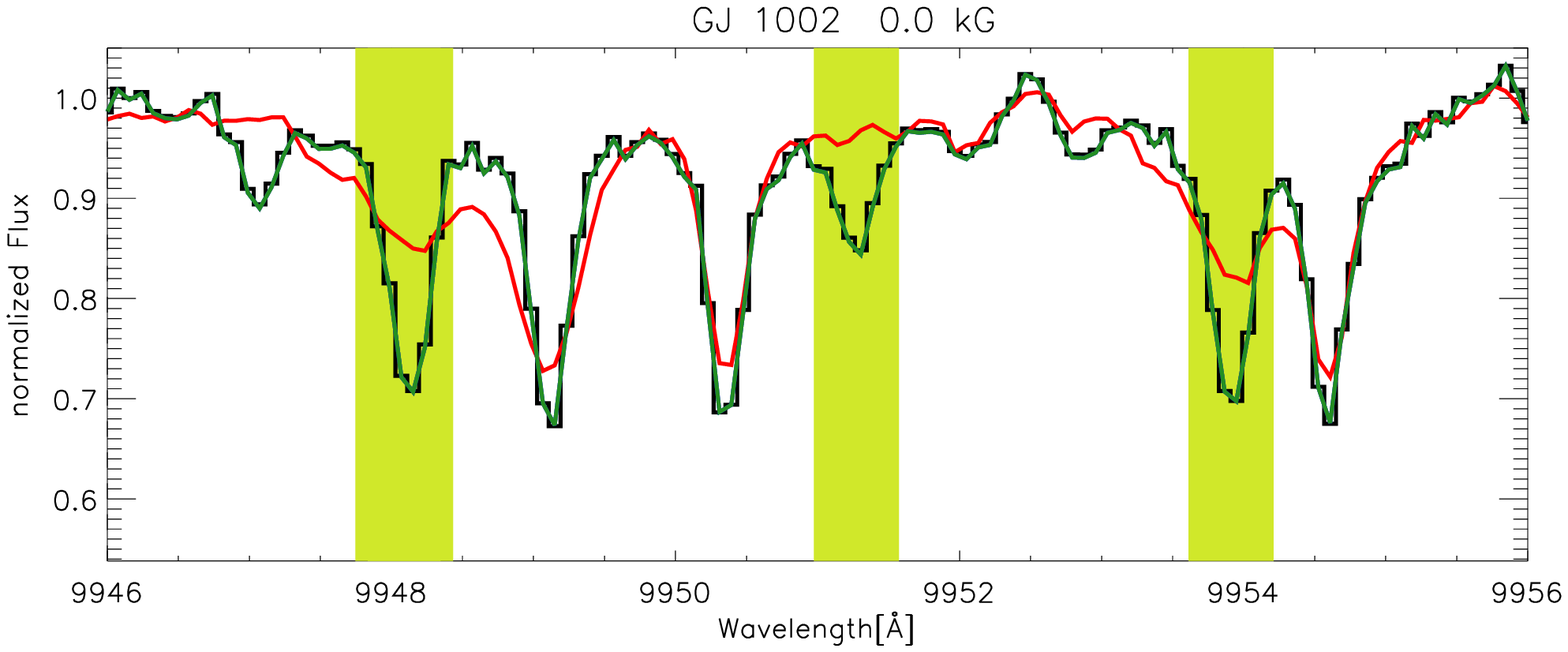}
  \caption{\label{fig:fit5}Same as Fig.\,\ref{fig:fit1} for GJ~1245AB (M5.5), GJ~1286 (M5.5), and GJ~1002 (M5.5).}
\end{figure}

\begin{figure}
  \includegraphics[width=\hsize,clip=,bbllx=0,bblly=190,bburx=648,bbury=468]{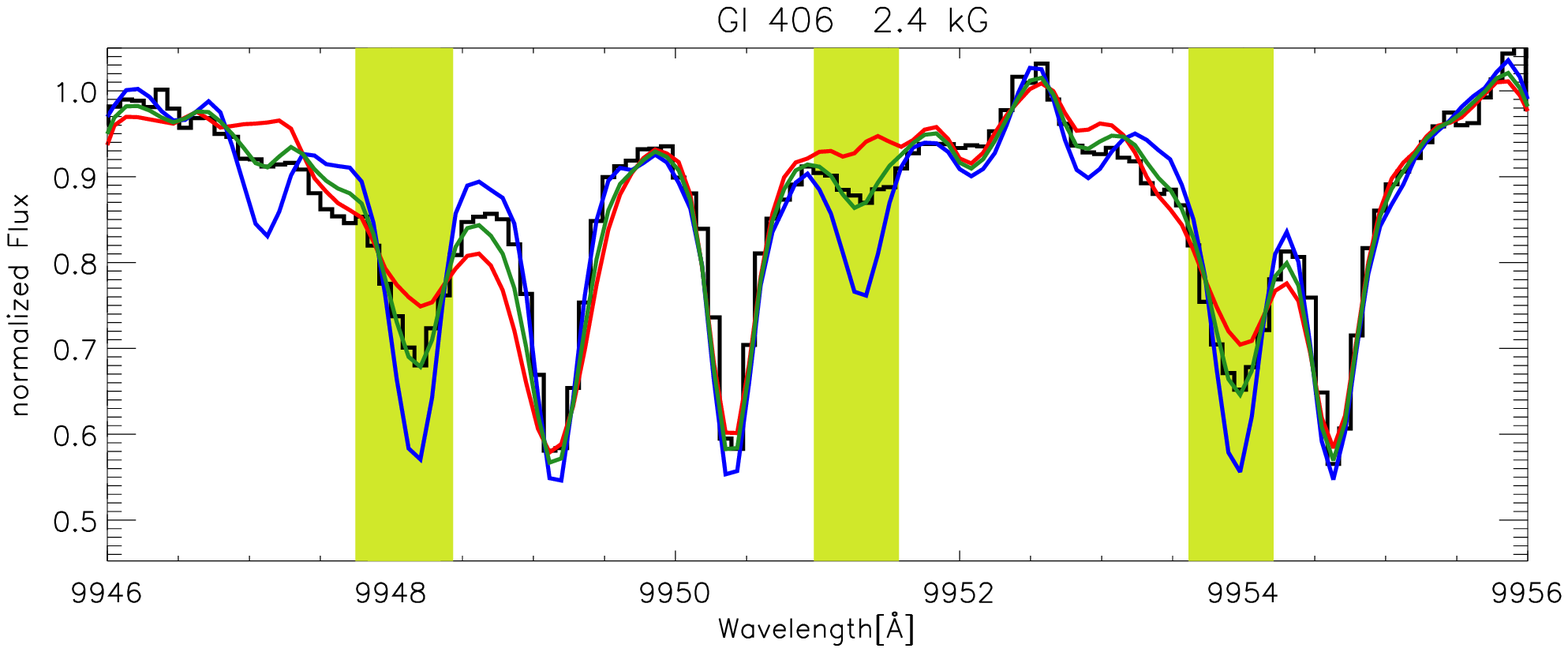}
  \includegraphics[width=\hsize,clip=,bbllx=0,bblly=190,bburx=648,bbury=468]{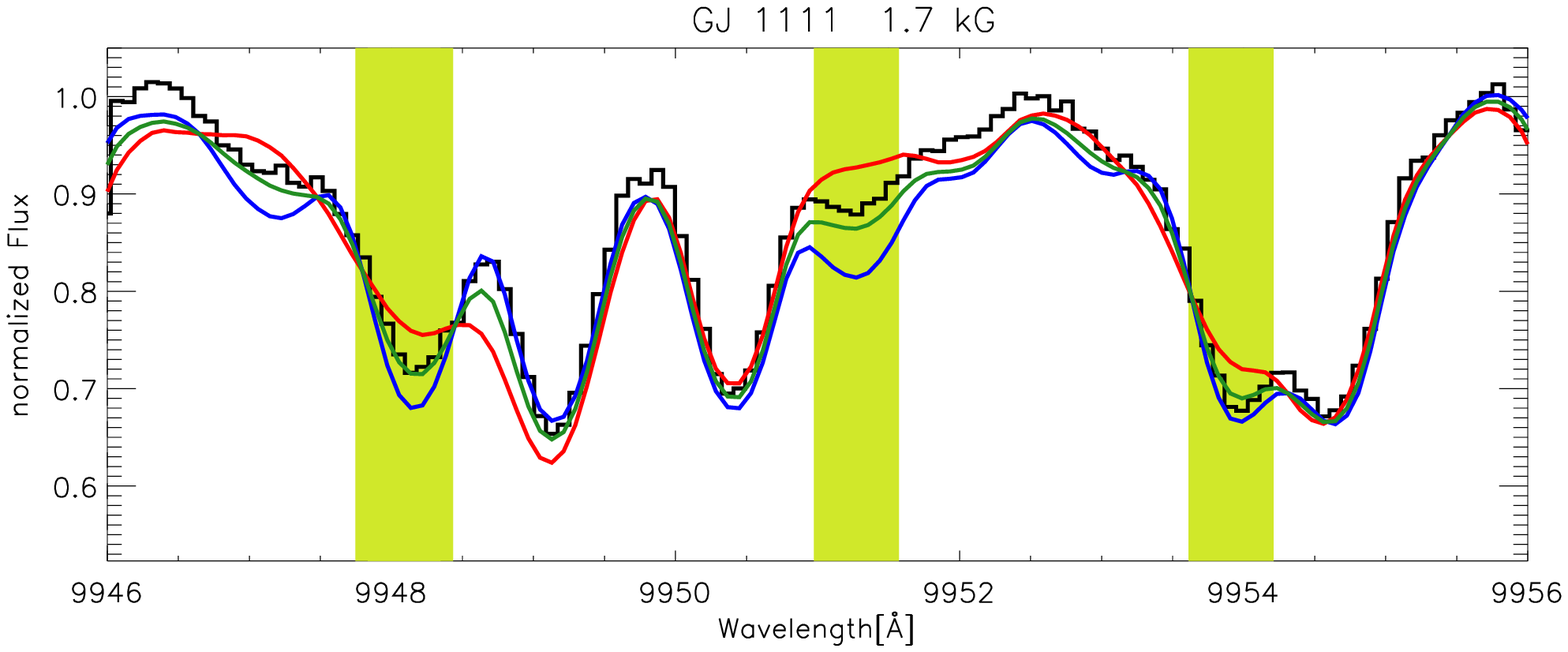}
  \includegraphics[width=\hsize,clip=,bbllx=0,bblly=190,bburx=648,bbury=468]{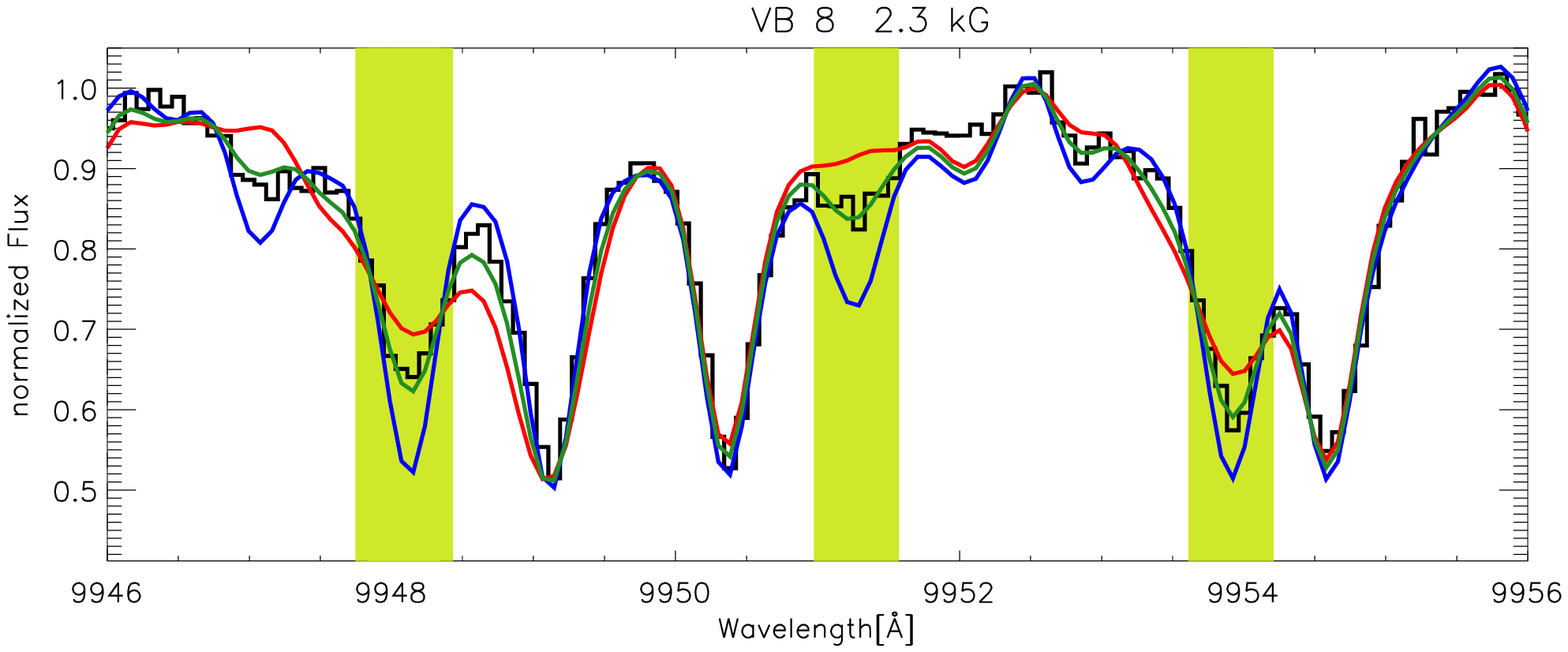}
  \caption{\label{fig:fit6}Same as Fig.\,\ref{fig:fit1} for Gl~406 (M5.5), GJ~1111 (M6), and VB~8 (M7)}
\end{figure}

\begin{figure}
  \includegraphics[width=\hsize,clip=,bbllx=0,bblly=190,bburx=648,bbury=468]{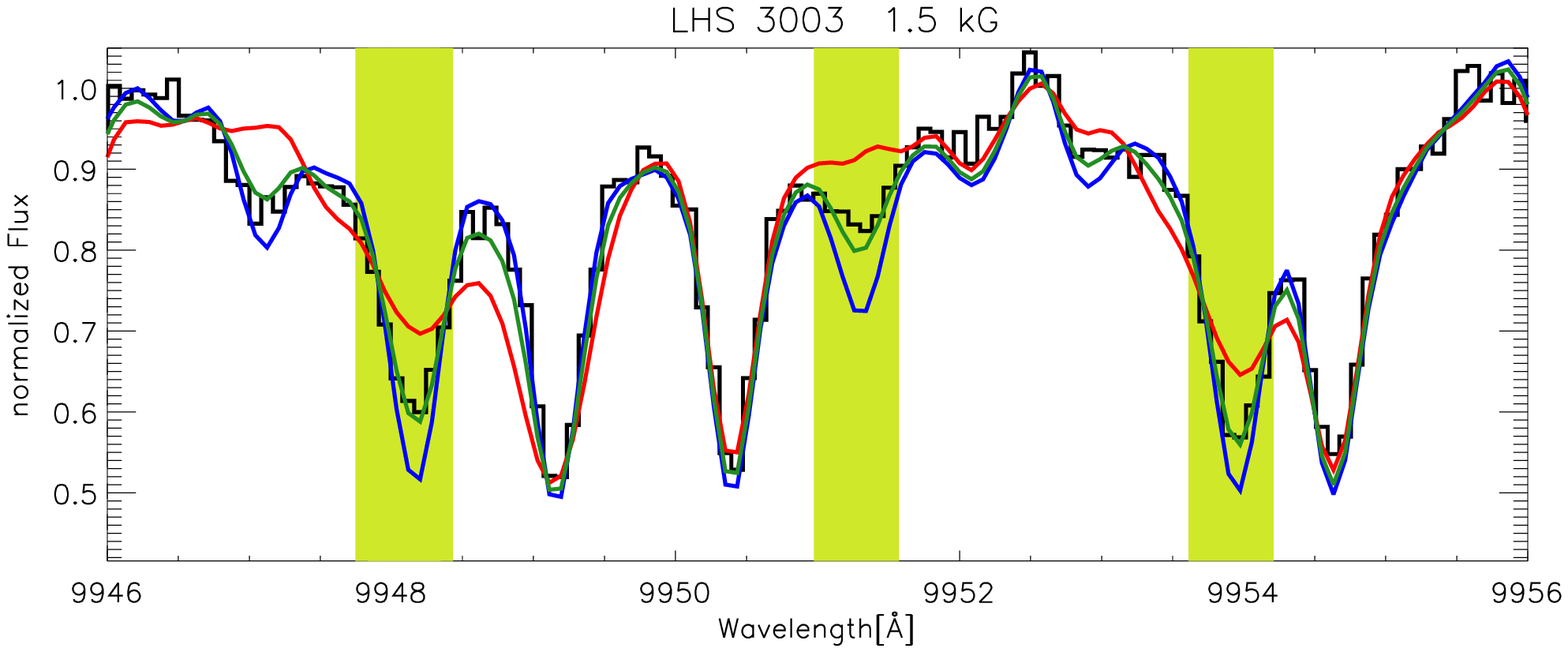}
  \includegraphics[width=\hsize,clip=,bbllx=0,bblly=190,bburx=648,bbury=468]{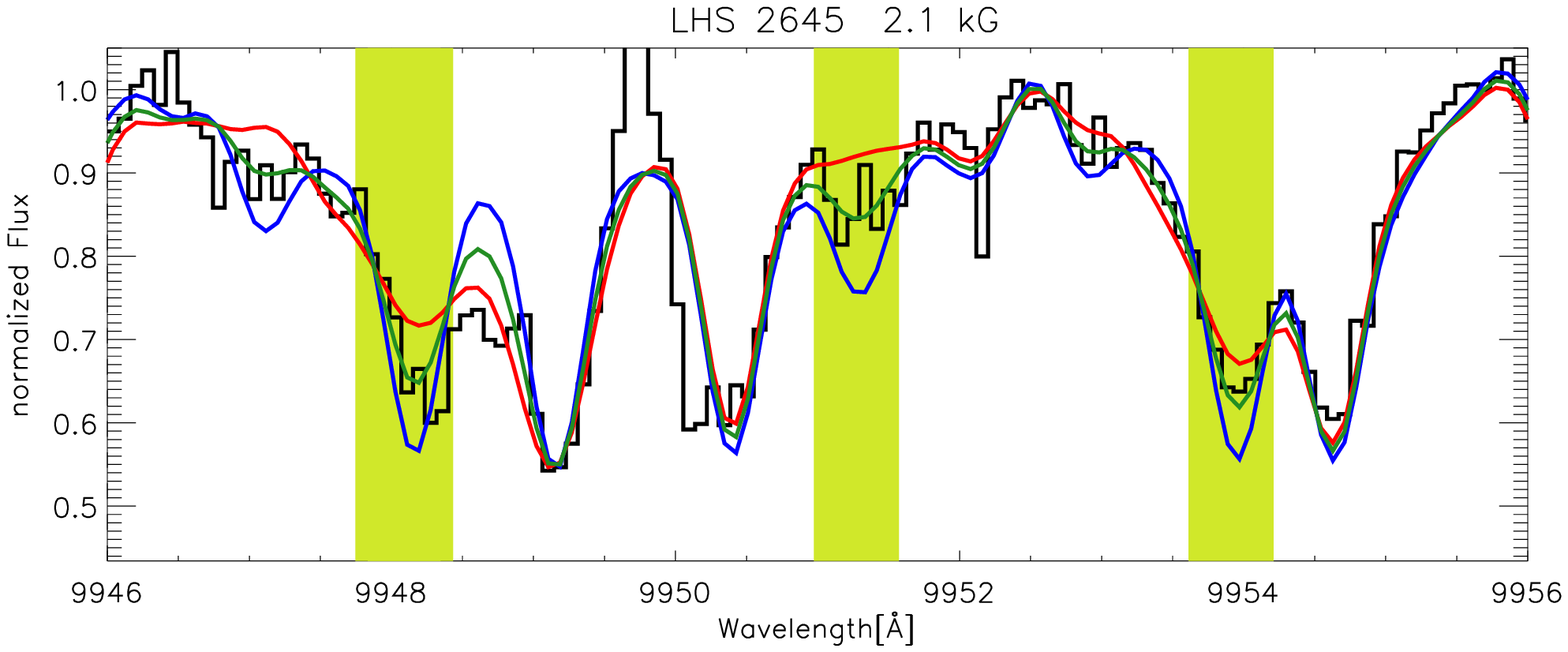}
  \includegraphics[width=\hsize,clip=,bbllx=0,bblly=190,bburx=648,bbury=468]{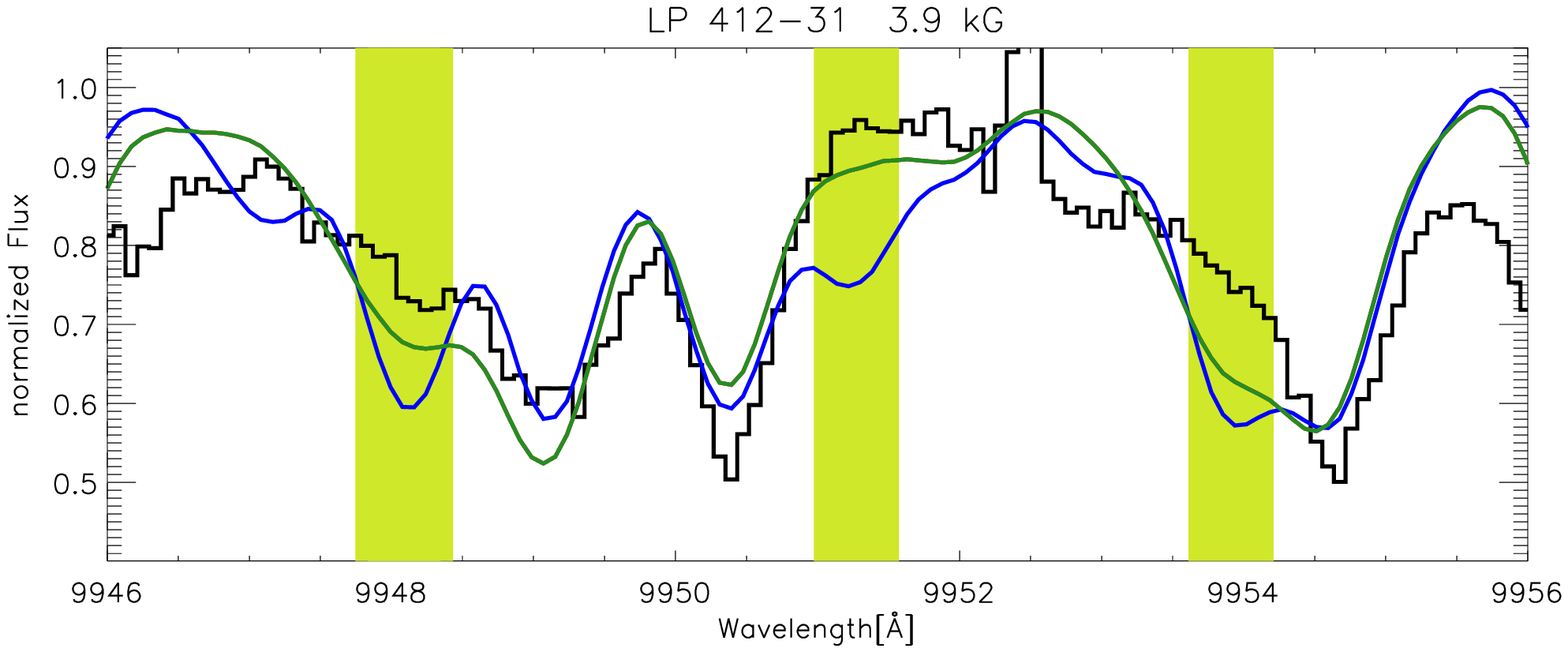}
  \caption{\label{fig:fit7}Same as Fig.\,\ref{fig:fit1} for LHS~3003 (M7), LHS~2645 (M7.5), and LP~412$-$31 (M8). There is no red line in the bottom panel because the green line is coincident with it.}
\end{figure}

\begin{figure}
  \includegraphics[width=\hsize,clip=,bbllx=0,bblly=190,bburx=648,bbury=468]{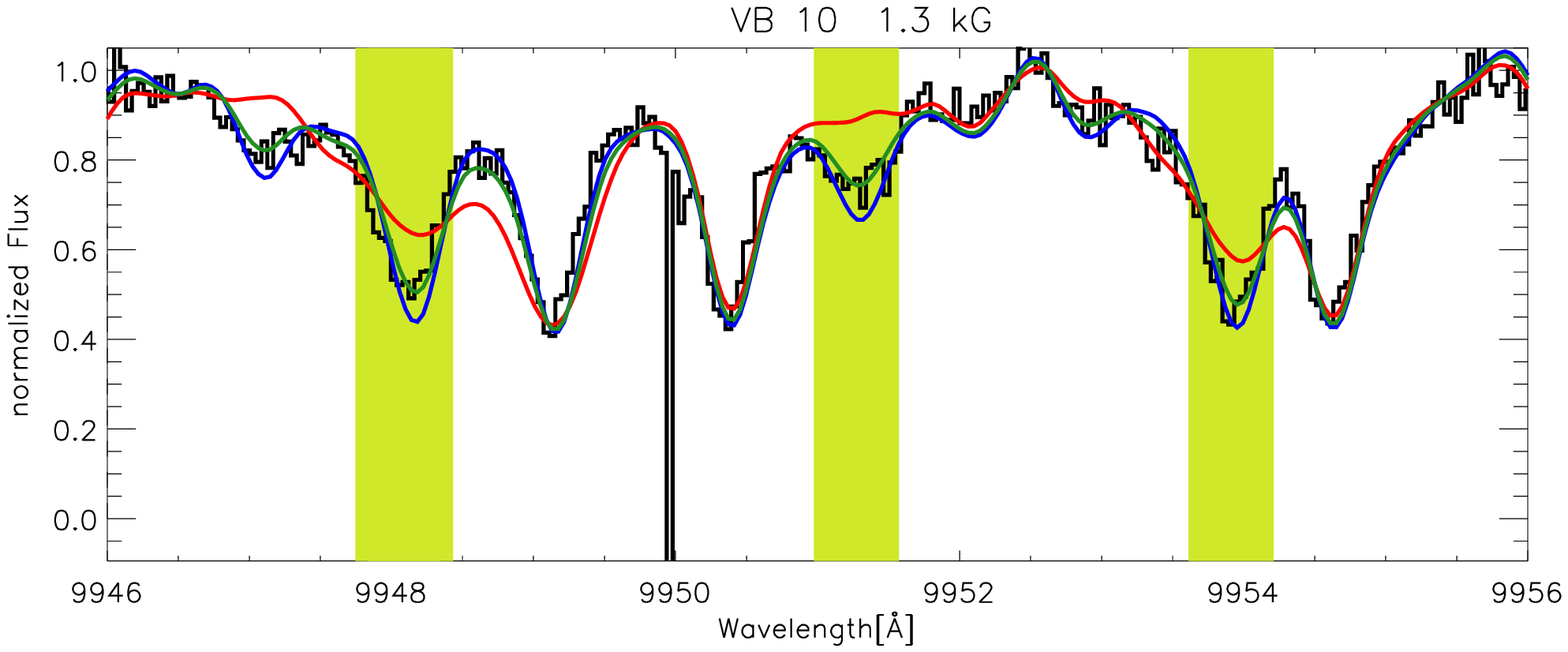}
  \includegraphics[width=\hsize,clip=,bbllx=0,bblly=190,bburx=648,bbury=468]{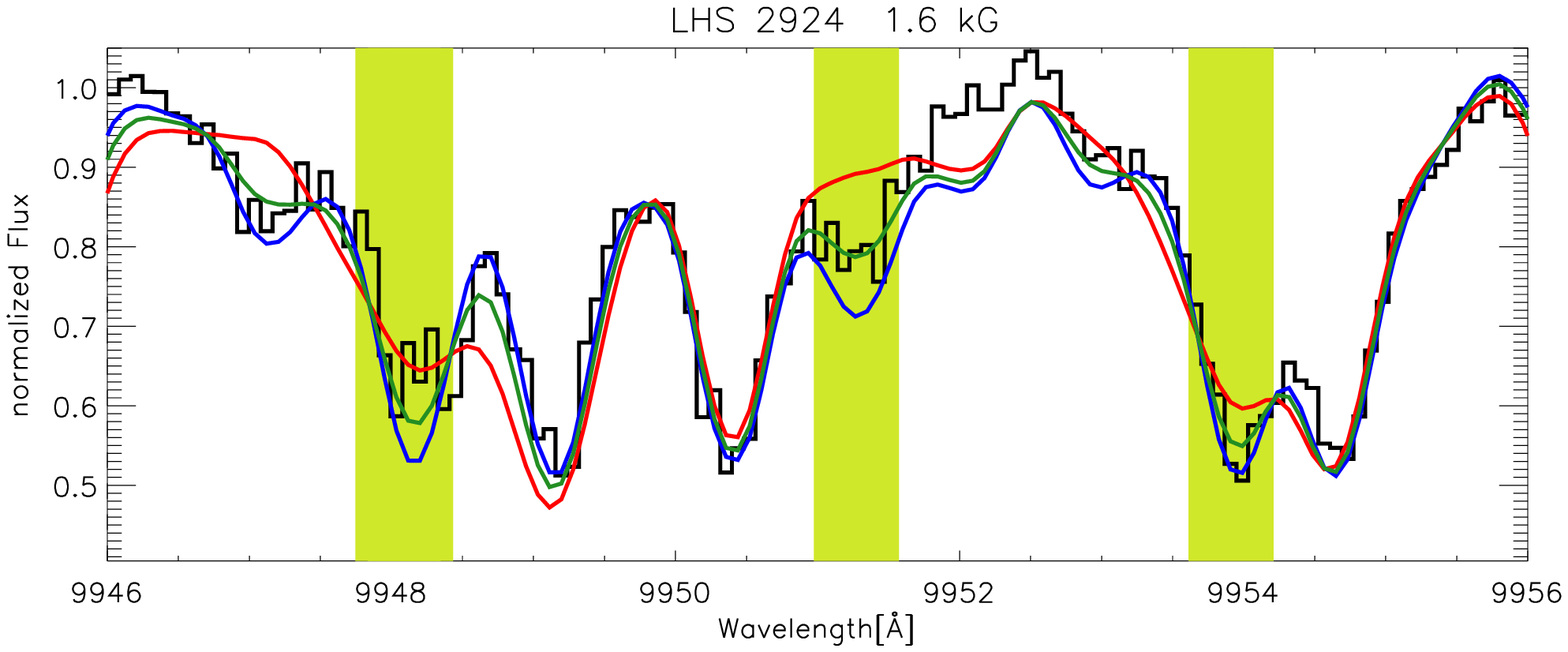}
  \includegraphics[width=\hsize,clip=,bbllx=0,bblly=190,bburx=648,bbury=468]{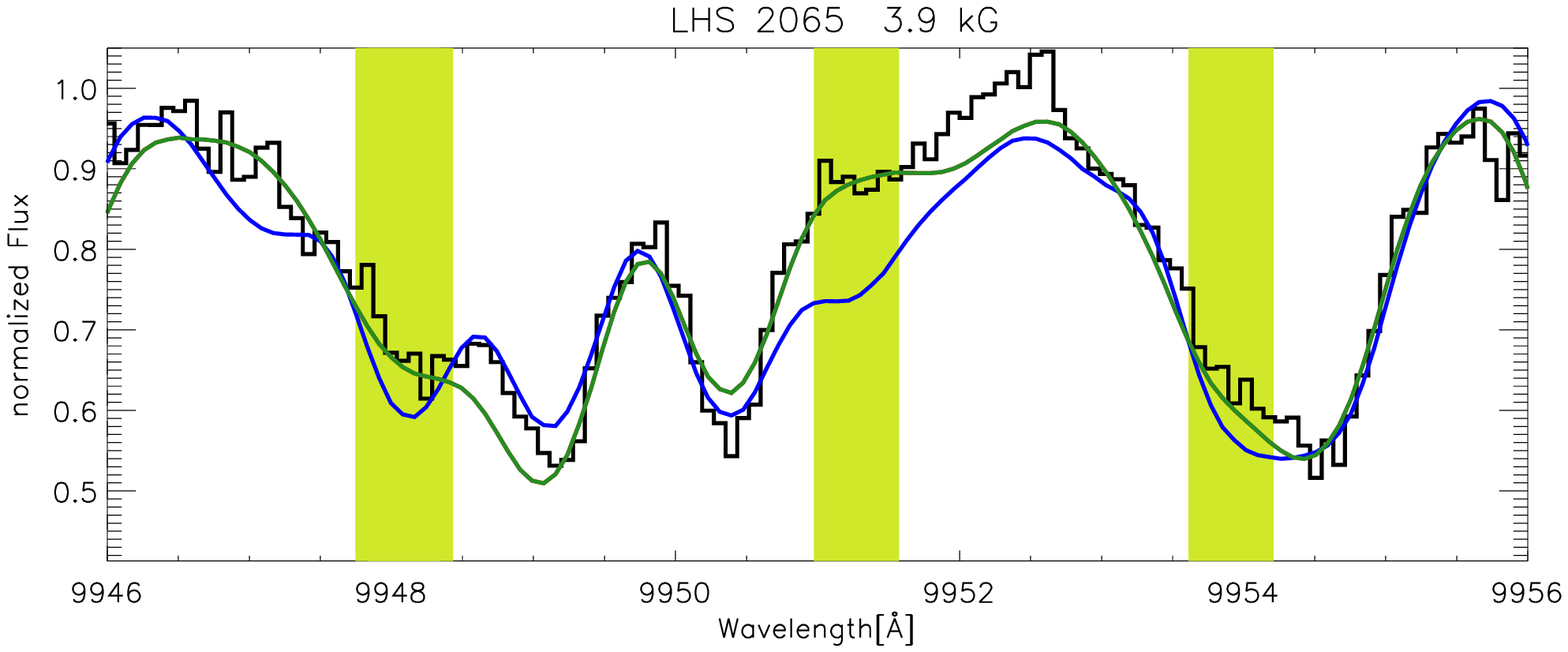}
  \caption{\label{fig:fit8}Same as Fig.\,\ref{fig:fit1} for VB~10 (M8), LHS~2924 (M9), and LHS~2065 (M9). There is no red line in the bottom panel because the green line is coincident with it.}
\end{figure}

The results of our spectral analysis are summarized in
Table\,\ref{tab:Results}. For all targets, projected rotation velocity
$v\,\sin{i}$, FeH strength $a$, and H$\alpha$ activity in terms of
log\,($L_{\rm H\alpha}/L_{\rm bol}$) are given. Literature values of
X-ray activity log\,($L_{\rm X}/L_{\rm bol}$) are included where
available.  We analyzed the spectra of 24 objects of spectral types
between M2 and M9; they cover the temperature region between 2400 and
3600\,K. The two stars Gl\,873 (M3.5) and GJ\,1002 (M5.5) were used as
reference objects. For each object, the wavelength region 9946 --
9956\,\AA, which is one of the four regions we used for the fitting
procedure, is shown in Figs.\,\ref{fig:fit1} -- \ref{fig:fit8}. In
these plots, data is plotted in black, the best fit is overplotted as
a green line.  Additionally, we show the pure enhanced and
rotationally broadened spectra of GJ\,1002 and Gl\,873, i.e. the two
bracketing extremes of zero field strength and a field of $Bf \approx
4$\,kG in blue and red color, respectively. Three lines that are
particularly sensitive to magnetic splitting are indicated with light
green bars. Here the signature of magnetic broadening is most
obvious. Several magnetically insensitive lines show how lines are
broadened without the influence of magnetic fields, i.e. the
broadening predominantly due to rotation.

In the spectrum of some of the warmer objects, (Gl\,70, AD\,Leo, Gl,
876, and GJ\,1005A), the Ti line at 9949\,\AA\ causes a difference
with the template spectra, so we excluded the Ti line from the fit. In
all cases, the quality of the fit in the magnetically insensitive
lines shows that the rotation velocity cannot be higher than in the
achieved fit, and the magnetically sensitive lines allow us to
reliably distinguish between weak or zero field strength, intermediate
field strength, and strong fields.

In two targets, LP\,412-31 and LHS\,2065, the best solution does not
accurately reproduce the target spectra, but instead appears to
require a stronger field than our high-field standard. In them,
rotational broadening is above 10\,km\,s$^{-1}$ and FeH absorption is
very strong, so that saturation and rotational broadening wash out the
structure of individual FeH lines. Furthermore, data quality in these
faint targets is not as high as in the earlier M-type stars. We show
the spectra of LP\,412-31 and LHS\,2065 in the lower panels of
Figs.\,\ref{fig:fit7} and \ref{fig:fit8}, respectively.  The best fits
come from the enhanced and rotationally broadened spectrum of Gl\,873
(i.e., $p = 1$ in Eq.\,\ref{eq:S}), but it does not convincingly match
the data in both spectra.  On the other hand, both spectra show a
behavior that is much closer to the strong-field case (the green line
sits on top of the red line) than to the weak-field case (blue)
especially in the magnetically sensitive regions. In the case of
LP\,412-31, the spectrum even seems to be a natural extrapolation of
the sequence of spectra from the blue (no field) over the green ($Bf
\approx 4$\,kG) spectrum towards the observed data. Based on this we
suggest that the magnetic flux $Bf$ in LP\,412-31 is substantially
higher than on Gl\,873, i.e. that the filling factor $f$ is larger
than on Gl\,873. The mean field in LHS\,2065 seems to be only slightly
stronger than the 4\,kG in Gl\,873. As a consistency check, we used
the method of line ratios as explained in Paper~I, where we measured
ratios of two neighbored lines, of which one is magnetically sensitive
and the other is magnetically insensitive. From the pair at
9948/9950\,\AA\ we derive ratios consistent with a minimum flux of
3.9\,kG in both targets. The values from the pairs at 9954/9955\,\AA\ 
exceed the valid range of that ratio and also suggest a field strength
in excess of the one in Gl\,873.

\placetable{tab:Results}

\subsection{Rotation velocities}

The rotation velocities of our objects are presented in the fourth
column of Table\,\ref{tab:Results}. We think that the rotation
velocities derived from the FeH band adjusted for spectral type may be
superior to those derived from comparison to warmer templates for two
reasons: (a) the modification of the FeH strength provides a template
that much better reproduces the unbroadened spectrum of VLMS than an 
uncorrected spectrum of a slowly rotating M-type dwarf does, and 
(b) the structural richness of the FeH band gives a clearer
signal than most other regions in the spectra do. This will be even
more important for the analysis of L-dwarfs.

For LHS\,3003 we are not aware of any determination of $v\,\sin{i}$ in
the literature. All other measurements are consistent with the values
we found in the literature except for AD Leo (discussed below). All
the objects later than M6 show detectable rotation. 

\begin{figure}
  \includegraphics[width=\hsize,clip=,bbllx=0,bblly=190,bburx=648,bbury=468]{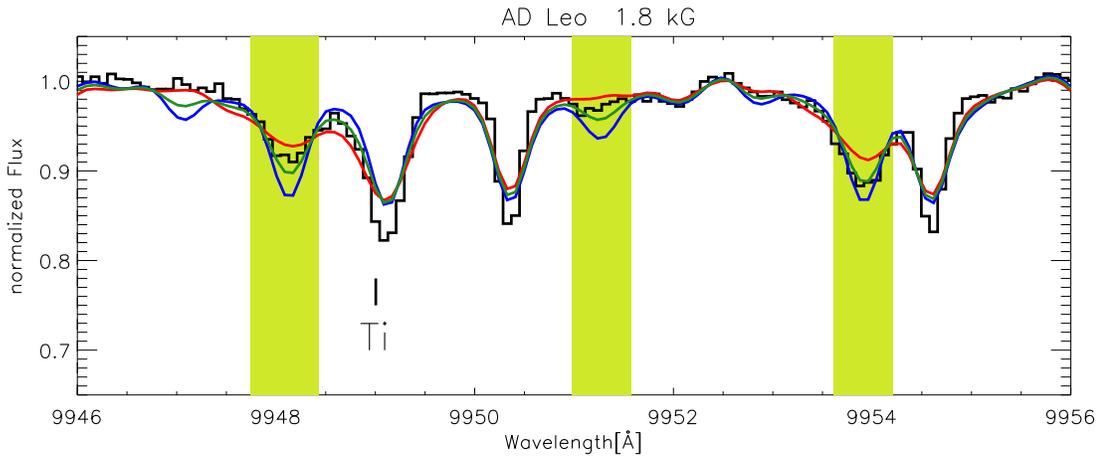}
  \caption{\label{fig:ADLeo6kms}Same as upper panel in Fig.\,\ref{fig:fit2} (AD~Leo, M3.5) but with $v\,\sin$ fixed at 6 km\,s$^{-1}$.}
\end{figure}

\cite{Vogt83, Marcy92, Delfosse98} and \cite{Fuhrmeister04}
spectroscopically measured $v\,\sin{i}$ in AD~Leo and all report
rotation of $v\,\sin{i} = 5-7$\,km\,s$^{-1}$. We show the best fit
that we can achieve with a value of $v\,\sin{i} = 6$\,km\,s$^{-1}$ in
Fig.\,\ref{fig:ADLeo6kms} in the same manner as in the previous plots.
For this case, the magnetic flux we derive is diminished (only 1.8~kG
instead of 2.9~kG) compared to our best fit. The fit quality, however,
is worse, particularly in the Zeeman-insensitive lines at $\lambda =
9950.3$ and 9954.6\,\AA, which are clearly too broad in the fit
compared to the data. We thus think that the projected rotation
velocity of AD~Leo is around 3\,km\,s$^{-1}$ rather than
6\,km\,s$^{-1}$. We obtain similar results from our own more careful
cross-correlation analysis in selected TiO regions. The rotation
period of AD Leo has been suggested to be 2.7 days from photometric
variability \citep{SH86}, which would produce an equatorial rotation
velocity of around 10 \,km\,s$^{-1}$ (and therefore suggest that we
are viewing it at a low inclination angle).

\placefigure{fig:ADLeo6kms}

\subsection{Magnetic fields}

We present our measurements of integrated magnetic flux in the last
column of Table\,\ref{tab:Results}. These results comprise the first
direct detections of magnetic fields in objects later than M4.5; they
suggest that magnetic fields are ubiquitous in late-type dwarfs at
least down to spectral type M9. The M8 object LP\,412-31 and the M9
object LHS\,2065 appear to exhibit magnetic flux $Bf$ that are
probably stronger than in any other of our targets, which otherwise
have values of $Bf$ less than or on the order of 3-4\,kG.

\begin{figure}
  \plottwo{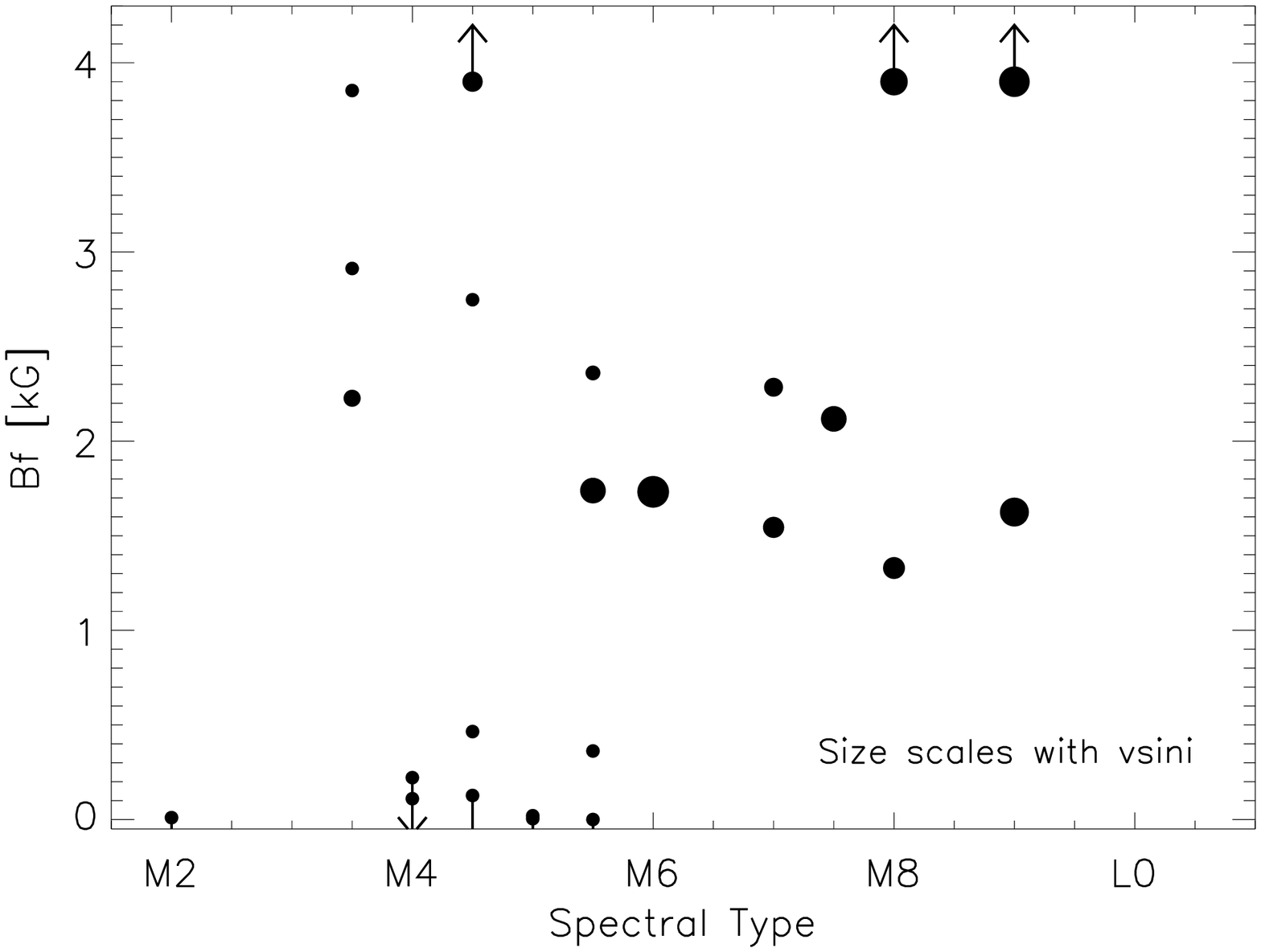}{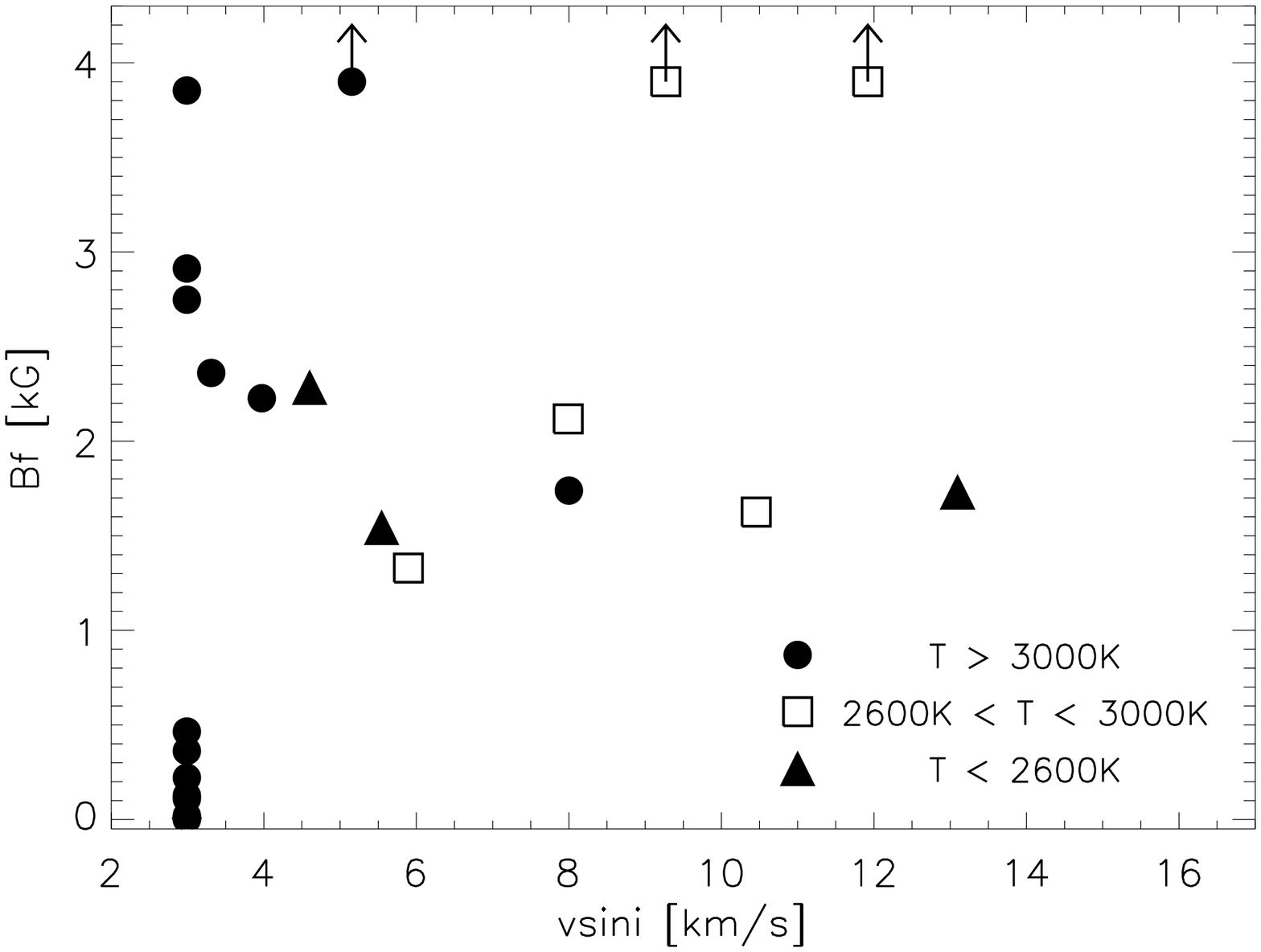}
  \caption{\label{fig:Bf}Left panel: Magnetic flux $Bf$ vs. spectral type; 
    the size of the symbols scales with projected rotation velocity
    $v\,\sin{i}$. Right panel: $Bf$ vs. $v\,\sin{i}$, three groups of
    effective temperature $T_{\rm eff}$ are distinguished by different
    symbols as indicated in the plot. In both panels, up- and downward
    arrows indicate upper and lower limits in $Bf$, respectively.}
\end{figure}

Our results on the magnetic fluxes are plotted as function of spectral
type and projected rotation velocity in the left and right panels of
Fig.\,\ref{fig:Bf}. In the left panel of Fig.\,\ref{fig:Bf} we
indicate the rotation velocity of the targets using large symbols for
(relatively) rapid rotators. In the right panel, stars cooler than
$T_{\rm eff} =2600$\,K are plotted as filled triangles, stars warmer
than $T_{\rm eff} =3000$\,K as filled circles, and intermediate stars
near to spectral type M7 are plotted as open squares.  For the
following, we call a mean field on the order of $Bf \sim 2$\,kG an
intermediate field, and a mean field on the order of $Bf \sim 4$\,kG a
strong field. Note that for solar-type stars, a flux of $Bf \sim
2$\,kG would be considered a strong field (even in K-type T Tauri
stars).

Both panels in Fig.\,\ref{fig:Bf} show that strong and intermediate
magnetic fields occur at the whole range of spectral types and
rotation velocities covered by our sample. In our early-type M dwarfs
(M2-3.5), the situation is not clear-cut. AD~Leo is a famous flare
star but doesn't show much rotational broadening. Gl~873 similarly has
strong H$\alpha$ emission and high magnetic flux but very low
rotational broadening. Gl~729 shows both rotation and a field; Gl~70
and Gl~876 show neither. This sample is too small, and should be
increased (although finding measurable rotators at this spectral type
is not easy). The total number of actual field measurements for the
(most easily observed) early M dwarfs is rather sparse (and confined
to dMe stars). Almost all of the sample in \citet{Delfosse98} are too
slowly rotating to show broadening, but that makes it easier to detect
field effects.  Clearly, more research should be conducted in this part
of parameter space.

Among the M4-5.5 stars there are 3 measurable rotators (although with
only a marginal detection in the case of Gl~406), all of which show
fields (the most active is the famous flare star YZ~CMi).  One slower
rotator also shows fairly strong flux and activity (GJ~1224); again
there is the possibility that inclination is making the rotational
broadening small in this case. Otherwise the 8 slow rotators have
no field detected. This appears to be reasonable evidence for a
rotational influence. It has already been noted by \cite{Mohanty03},
though using proxies instead of actual flux measurements.

We find no clear relation between rotation velocity and mean magnetic
field in our late-M sample. Instead, all cool objects exhibit
detectable magnetic flux. No rotation-activity relation was found in
very late objects \citep{Mohanty03}, so we did not expect a dependence
between $v\,\sin{i}$ and $Bf$ in objects later than $\sim$\,M7.
However, the two cases in which we detect a very strong mean field are
rapid rotators with $v\,\sin{i} > 10$\,km\,s$^{-1}$. This could
indicate a dependence of field generation on rotation at late spectral
types of some sort, but such a suggestion is hampered by the lack of
slowly rotating late-M objects in general (which would help define
it), and certainly is not statistically significant in our sample.  It
is also worth noting that the filling factor is thought to be near
unity for dMe stars (or T Tauri stars), so it is likely that the field
strength itself is greater in these very strong objects.

\cite{Noyes84} pointed out the potential relevance of Rossby number
($P_{rot} / \tau_c$) to the production of magnetic fields in
convective dynamos, where $\tau_c$ is the convective overturn
timescale and $P_{rot}$ is the rotational period. \cite{Mohanty03} did
not find strong evidence that it matters for fully convective stars.
Whether the Rossby number is relevant or not depends in part on the
length scale at which the field is produced, and whether it is large
enough to let rotation play a role. We now have a chance to further
probe this question. A crude estimate for $\tau_c$ in fully convective
stars can be derived as follows. An expression for convective
velocities in mixing length theory is ${v_c}^3 \propto ({{\Gamma -
    1}\over \Gamma}) {L_* \over {{R_*}^2 \rho}}$ \citep{Thom93}.
Since these objects are fully convective, we can take the convective
luminosity to be the stellar luminosity, the radius to be the stellar
radius, and the density to be the mean stellar density.  For very
low-mass stars, radius scales like mass.  Luminosity is often assumed
to behave like $M^{\beta}$.  Then we find ${v_c}^3 \propto {{L_* R_*}
  \over M_*}\rm{\ \ \ or\ \ \ } v_c \propto {M_*}^{\beta /3}.$ The
first of these two results can also be obtained from a dimensional
analysis.  Since the convective overturn time can likewise be
approximated as $\tau_c \propto {R_* / v_c}$; we find that $\tau_c
\propto {M_*}^{1 - (\beta /3)}$.

Often $\beta$ is taken to be roughly 5 for this mass range, in which
case $\tau_c$ is inversely proportional to ${M_*}^{(2/3)}$. The
situation is actually somewhat more complicated (even if we simplify
by restricting ourselves to talking about objects several Gyr old).
To estimate $\beta$ we used the models of \citet{ChaBar00} and looked
at gradients in luminosity over small increments of mass.  Above and
near the fully convective boundary $\beta \approx 3$ (which makes
$\tau_c$ almost constant), but then it becomes very steep towards the
substellar boundary, reaching values of several tens. This means that
the convective overturn times grow quite long near the substellar
boundary (where degeneracy is taken over from fusion). That is
exaggerated further by the fact that radius is no longer linear in
mass, as degeneracy causes it to become nearly constant (then depend
inversely on mass).  The rotation periods also vary greatly over the
mass range of interest (from weeks for inactive early-M stars to days
for active mid-M stars to hours for late-M stars). Rotation must
therefore be the driving force behind any dependence on Rossby number
for stars above about two-tenths of a solar mass, but convection is
probably more important at the bottom of the main sequence.  What is
clear is that both the longer convective overturn times and
increasingly rapid rotations of the very cool objects push the Rossby
numbers to very small values. This obviously does not hamper the
generation of strong, widespread fields, however.  That is a valuable
clue for the study of fully convective dynamos.

\placefigure{fig:Bf}

\subsection{H$\alpha$ activity and magnetic flux}

\begin{figure}
  \plottwo{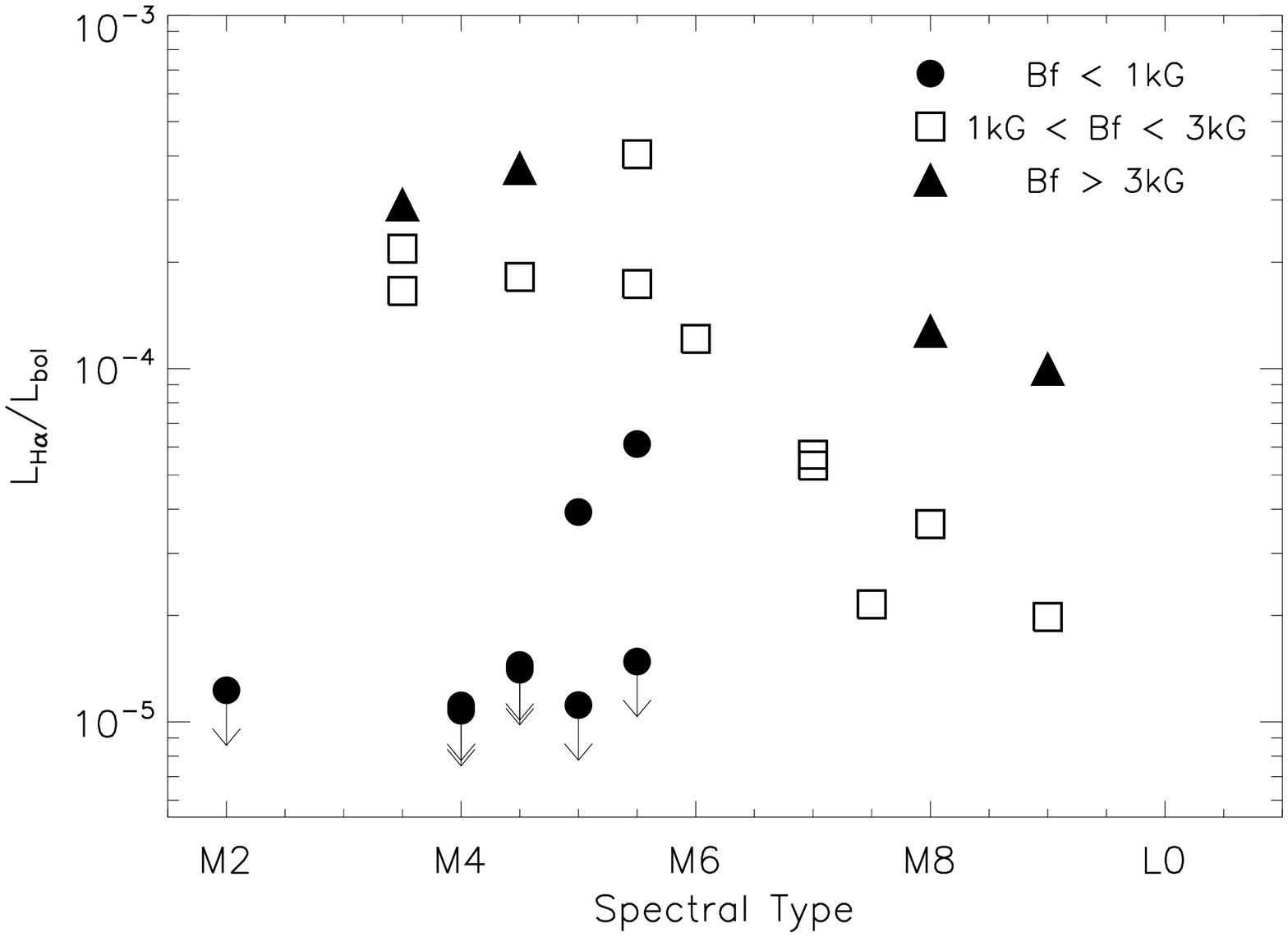}{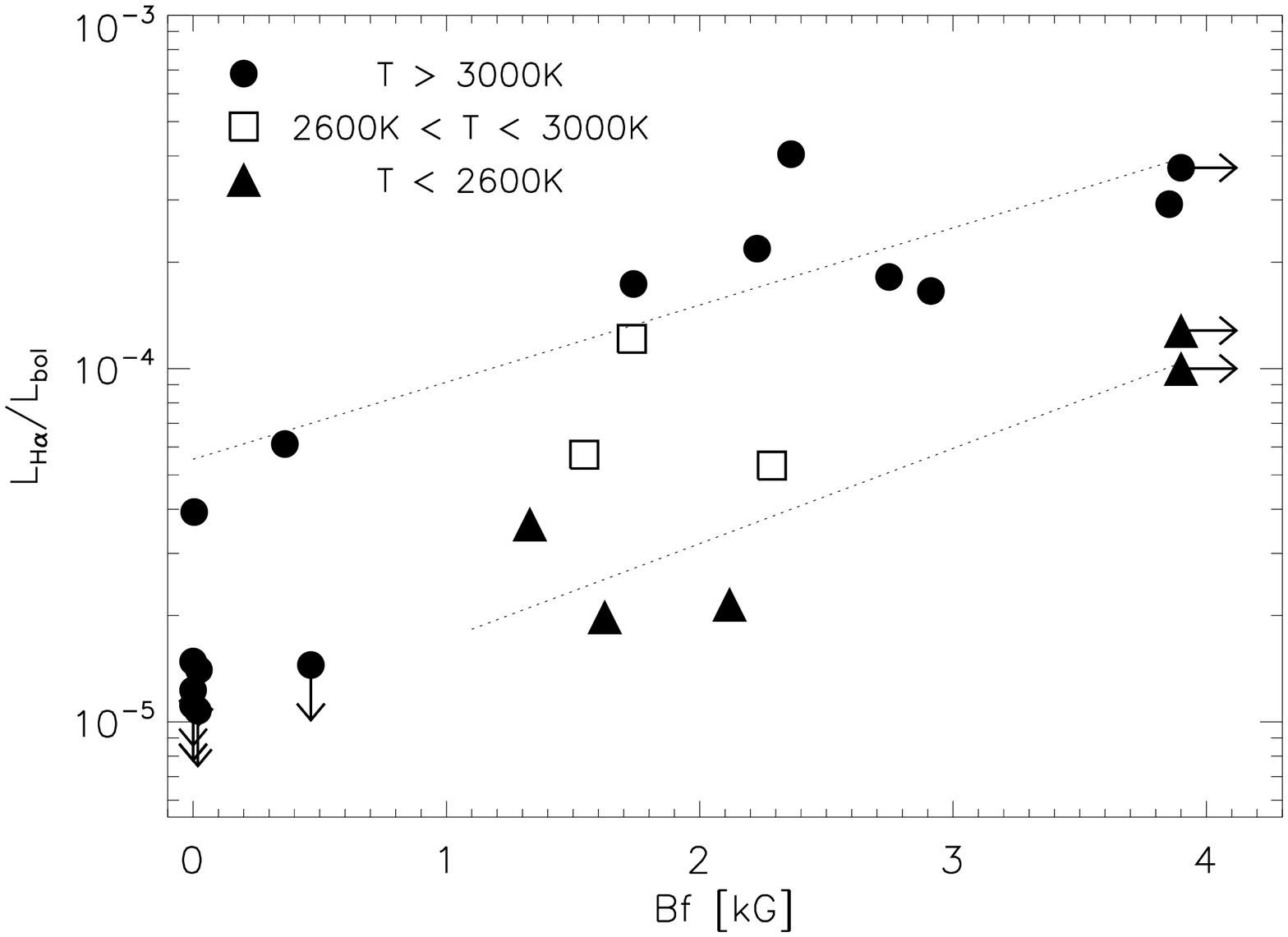}
  \caption{\label{fig:Lalpha}Normalized H$\alpha$ luminosity 
    log\,$L_{\rm H\alpha}/L_{\rm bol}$. Left panel: log\,$L_{\rm
      H\alpha}/L_{\rm bol}$ vs. spectral type, symbols distinguish
    objects with weak or no magnetic flux, intermediate flux, and
    strong flux as shown in the plot. Right panel: log\,$L_{\rm
      H\alpha}/L_{\rm bol}$ vs. $Bf$. Symbols indicate temperature as
    in the right panel of Fig.\,\ref{fig:Bf}.}
\end{figure}

Our measurements of activity from H$\alpha$ and of the magnetic flux
are derived from the same spectrum. We can thus directly investigate
the dependence of H$\alpha$ activity on the magnetic field even though
both may well be variabler. We plot our measurements of activity in
terms of the ratio of the luminosity in H$\alpha$ to the bolometric
luminosity, log\,$L_{\rm H\alpha}/L_{\rm bol}$, as a function of
spectral type in the left panel of Fig.\,\ref{fig:Lalpha}. Symbols
indicate magnetic flux as explained in the caption.  This plot
resembles Fig.\,7 from \cite{Mohanty03} and confirms their findings;
in the stars earlier than M7, activity measurements span a range of at
least two magnitudes with several non-detections. All objects of
spectral types M6~--~M9 show significant H$\alpha$ emission (no lower
limits), and a steep fall-off in log\,$L_{\rm H\alpha}/L_{\rm bol}$ is
observed towards the coolest objects as in \cite{Gizis00} and 
\cite{West04}. Among the active stars only, the level of activity 
is continually decreasing towards later spectral type.

In the same plot, we indicate magnetic field strengths by different
symbols. For each spectral type, the highest value of log\,$L_{\rm
  H\alpha}/L_{\rm bol}$ is measured in the stars with the strongest
fields, and all inactive stars have weak (or no) magnetic fields. This
means that at given spectral type log\,$L_{\rm H\alpha}/L_{\rm bol}$
correlates with magnetic flux. The plot of log\,$L_{\rm
  H\alpha}/L_{\rm bol}$ as a function of magnetic flux $Bf$ in the
right panel of Fig.\,\ref{fig:Lalpha} demonstrates this relation (we
use the same symbols as before in the right panel of
Fig.\,\ref{fig:Bf}).

Among the two groups of stars that have temperatures warmer than
3000\,K (full circles) and cooler than 2600\,K (full triangles), the
measured level of activity is proportional to the magnetic flux $Bf$
among the stars that show H$\alpha$ emission. All non-active stars
have weak or zero field strength. It can also be seen that activity
drops towards later spectral type, i.e. there is an offset between the
two groups.  Three targets are of intermediate temperature (spectral
type $\sim$\,M7), and they fall between the two groups of warm and
cool objects. We describe the relations among the warm and the cool
objects by independently fitting a linear model to the active stars in
these groups.  The relations are overplotted in the right panel of
Fig.\,\ref{fig:Lalpha} as dotted lines, they are best approximated by
the following parameterizations:

\begin{eqnarray}
  \label{eq:Ha_Bf}
  \mathrm{earlier~than~M7} & : & \mathrm{log} \frac{L_{\mathrm H\alpha}}{L_\mathrm{bol}} = -4.26 + 0.22 Bf\\ 
  \mathrm{later~than~M7} & : & \mathrm{log} \frac{L_{\mathrm H\alpha}}{L_\mathrm{bol}} = -5.03 + 0.27 Bf
\end{eqnarray}

If we assume an uncertainty in the measurement of the magnetic flux of
$\Delta Bf = 1$\,kG, we formally derive correlation coefficients of
$(0.22 \pm 0.07)$ and $(0.27 \pm 0.16)$ for the two subsamples. The
dependence of log\,$L_{\rm H\alpha}/L_{\rm bol}$ on $Bf$ appears to be
similar in both groups, although a flatter slope for the cooler group
might result if we had an actual value for LP\,412-31. The fact that
the three objects of intermediate temperature fall between the two
relations indicates a continuous dependence of $Bf$ on spectral type
rather than an abrupt change at spectral type M7.

\placefigure{fig:Lalpha}

All the objects M6 or later show detectable rotation and also
H$\alpha$ emission.  Those with rotations similar to the M4.5-5.5
rotators of 5-10\,km\,s$^{-1}$ (namely VB~8 and 10, LHS~3003 and 2645,
GJ~1002) have intermediate fluxes (the comparable mid-M stars are
intermediate or strong), while those above 10\,km\,s$^{-1}$ have very
strong fields (LP~412-31 and LHS~2065). Even in these two cases,
however, the H$\alpha$ emission is not as strong as for any of the
mid-M rotators. One star rotates pretty rapidly and has an
intermediate flux, but does not have as strong activity (GJ~1111); it
is at the warm end of this subset (M6).  This may mark the beginning
of the weakening of the field-activity connection with falling
photospheric temperature. That is the effect predicted to result from
increasing atmospheric neutrality (which decouples the field from
atmospheric motions), although it is at a little warmer temperature
than models suggest. This is consistent with what was found by
\cite{West04}, who also found that not all late-M stars showed
H$\alpha$ emission. It would be good to try to observe a few such
inactive stars to determine whether they are also slow rotators, 
and do not show detectable fields.

There appears to be no reason to imagine that the L dwarfs (at least
the early-type ones) are inactive due to an inability to generate
fields. That is consistent with the inference of fields on L dwarfs by
\citep{Berger05, Berger06}. Our suggested trend of increasing field at
later spectral types, however, is somewhat opposite of the impression
one gets from Fig.\,2 of \cite{Berger06}, although in both cases the
samples of field detections are too small for clear conclusions. We
note that our field detections are much more direct than those
inferred from gyrosynchrotron radio emission by \cite{Berger06}; ours
refer directly to the stellar surface rather than a radio-emitting
region well above the star and require no assumptions about
electron-density power spectrum, or emitting-region geometry. We are
therefore able to conclude that rather strong fields exist in the
photosphere, contrary to the tentative conclusion those authors make
that the fields may be somewhat weaker than in earlier M dwarfs.

\section{Conclusions}

This is the first attempt to measure magnetic fields directly on stars
with effective temperatures below 3200K. At these temperatures the
objects are firmly in the fully convective regime, so one is studying
dynamo action of the sort that is probably most common in stars and
disks (the cyclical solar dynamo is thought to operate only at
radiative-convective interfaces). We find evidence that the
increasingly small Rossby numbers near the bottom of the main sequence
are no hindrance to the generation of magnetic flux.

The sample of stars in this study is too small and biased to draw
general conclusions from it. We offer several conjectures supported by
our results that can be followed up by more work using our new
methodology.  These are: 1) magnetic fields exist, and can be quite
strong, throughout the M spectral sequence; 2) there may be a
connection between the rotation of the star and the amount of flux it
produces, although it is not yet clearly known what it is (and how
saturation plays a role in it); 3) there is a correlation between the
magnetic flux and the fractional amount of H$\alpha$ emission observed
when such emission is seen; 4) there is evidence that the fractional
amount of H$\alpha$ emission falls with photospheric temperatures in
late-M dwarfs regardless of how much magnetic flux they have,
supporting the proposition that activity fades at the bottom of the
M-sequence due to atmospheric neutrality rather than a fall in
magnetic flux production.

One of the most pressing questions left unanswered by this initial
work is whether we can actually measure fields on early L-dwarfs even
though they rarely exhibit emission activity. These objects are more
challenging for our method because of the heavy saturation of the FeH
lines, and the great tendency for rapid rotation in these objects.
The latter problem is almost certainly due to a lack of magnetic
braking, but it remains unresolved whether it is the field itself or
its coupling to the upper atmosphere that is to blame. Another current
mystery is why very young objects in the late-M spectral range appear
to be substantially more active than older objects of the same
temperature. That is a surprising result if the dominant effect
turning activity off in the very low-mass field objects is atmospheric
neutrality (which depends on temperature). Is this difference in
activity due to different magnetic flux generation at different ages,
or some other characteristic (eg. low gravity or primordial fields) in
younger objects? This question is more easily attacked with our
current methodology, and we have already begun observations to address it.

\acknowledgments It is a pleasure to thank J. Valenti, S. Solanki and
S. Berdyugina for enlightening discussions on the magnetic sensitivity
of the FeH molecule. We especially thank J. Valenti for his help with
the line data, and his very helpful comments as referee. 
We would also like to thank Matthew Browning for
helping us with the understanding of how dynamo theory can be informed
by these observations, and the derivation of convective overturn
timescales. This work is based on observations obtained from the W.M.
Keck Observatory, which is operated as a scientific partnership among
the California Institute of Technology, the University of California
and the National Aeronautics and Space Administration. We would like
to acknowledge the great cultural significance of Mauna Kea for native
Hawaiians and express our gratitude for permission to observe from
atop this mountain. GB thanks the NSF for grant support through
AST00-98468. AR has received research funding from the European
Commission's Sixth Framework Programme as an Outgoing International
Fellow (MOIF-CT-2004-002544).







\clearpage

\begin{deluxetable}{lcccc}
  \tablecaption{\label{tab:observations} Table of targets, exposure times and SNR measured in the FeH band}
  \tablewidth{0pt}
  \tablehead{\colhead{Name} & \colhead{SpT} & \colhead{$J$ mag} & \colhead{exposure [s]} & \colhead{SNR} }
  \startdata
  Gl 70              & M2.0  & \phn 7.37 & \phn 300 & 150\\
  Gl 729             & M3.5  & \phn 6.22 & \phn 120 & 145\\
  Gl 873             & M3.5  & \phn 6.11 & \phn 120 & 140\\
  AD Leo             & M3.5  & \phn 5.45 & \phn 120 & 100\\ 
  Gl 876             & M4.0  & \phn 5.93 & \phn 120 & \phn 85\\
  GJ 1005A           & M4.0  & \phn 7.22 & \phn 240 & \phn 90\\ 
  Gl 299             & M4.5  & \phn 8.42 &     1200 & \phn 65\\
  GJ 1227            & M4.5  & \phn 8.64 &     1200 & \phn 75\\
  GJ 1224            & M4.5  & \phn 8.64 & \phn 600 & \phn 75\\
  YZ Cmi             & M4.5  & \phn 6.58 & \phn 600 & \phn 70\\
  Gl 905             & M5.0  & \phn 6.88 & \phn 180 & \phn 60\\
  GJ 1057            & M5.0  & \phn 8.78 & \phn 300 & \phn 60\\
  GJ 1245B           & M5.5  & \phn 8.28 & \phn 300 & \phn 70\\
  GJ 1286            & M5.5  & \phn 9.14 & \phn 600 & \phn 50\\
  GJ 1002            & M5.5  & \phn 8.32 & \phn 300 & \phn 65\\
  Gl 406             & M5.5  & \phn 7.09 & \phn 200 & \phn 60\\
  GJ 1111            & M6.0  & \phn 8.24 & \phn 600 & \phn 50\\
  VB 8               & M7.0  & \phn 9.78 & \phn 600 & \phn 40\\
  LHS 3003           & M7.0  & \phn 9.97 & \phn 600 & \phn 35\\
  LHS 2645           & M7.5  &     12.19 &     2000 & \phn 30\\
  LP 412$-$31        & M8.0  &     11.76 &     2400 & \phn 30\\
  VB 10              & M8.0  & \phn 9.91 &     1100 & \phn 25\\
  LHS 2924           & M9.0  &     11.99 &     1800 & \phn 20\\
  LHS 2065           & M9.0  &     11.21 &     1200 & \phn 35\\
  \enddata
\end{deluxetable}

\begin{deluxetable}{ccc}
  \tablecaption{\label{tab:mag_regions}Spectral regions used for the fit}
  \tablewidth{0pt}
  \tablehead{\multicolumn{3}{c}{Slow rotators}}
  \startdata
  9895.5  & -- &      9905.5\,\AA \\
  9937.5  & -- &      9941.0\,\AA \\
  9946.0  & -- &      9956.0\,\AA \\
  9974.5  & -- &      9978.0\,\AA \\
  \noalign{\smallskip}
  \hline
  \noalign{\smallskip}
  \multicolumn{3}{c}{Rapid rotators}\\
  \noalign{\smallskip}
  \hline
  \noalign{\smallskip}
  9920.0  & -- &      9926.0\,\AA \\
  9932.0  & -- &      9937.0\,\AA \\
  9946.0  & -- &      9952.0\,\AA \\ 
  9977.5  & -- &      9982.0\,\AA \\
  \enddata
\end{deluxetable}

\begin{deluxetable}{lccrcrrrc}
  \tablecaption{\label{tab:Results} Results from our analysis. Effective temperature is from the calibration of \citet{Golimowski04}, X-ray luminosities are from the literature as indicated. We give formal 2$\sigma$ uncertainties for $Bf$, but we note that total uncertainties are on the order of a kilo-Gauss (see text).}
  \tablewidth{0pt}
  \tablehead{\colhead{Name} & \colhead{Sp} & \colhead{$T_{\rm eff}$} & \colhead{$v\,\sin{i}$} & \colhead{FeH $a$} &\colhead{H$\alpha$ EqW} & \colhead{log($\frac{L_\mathrm{H\alpha}}{L_\mathrm{bol}}$)} & \colhead{log($\frac{L_\mathrm{X}}{L_\mathrm{bol}}$)}  & \colhead{$Bf_{\rm fit}$}\\
    && [K] & [km\,s$^{-1}$] & &[\AA]&&& [kG]}
  \startdata
               Gl~70 & M2.0 &  3580 & $\le$3     &  0.5 & $ < 0.20$ &  $< -4.91$ & $<-4.44$\tablenotemark{a} &        $<  0.1$ \\
              Gl~729 & M3.5 &  3410 &    4       &  1.1 & $   4.47$ &  $  -3.66$ & $ -3.50$\tablenotemark{b} &   $ 2.2\pm 0.1$ \\
              Gl~873 & M3.5 &  3410 & $\le$3     &  0.9 & $   5.98$ &  $  -3.53$ & $ -3.07$\tablenotemark{a} &        $   3.9$ \\
              AD~Leo & M3.5 &  3410 & $\approx$3 &  0.8 & $   3.40$ &  $  -3.78$ & $ -3.02$\tablenotemark{a} &   $ 2.9\pm 0.2$ \\
              Gl~876 & M4.0 &  3360 & $\le$3     &  0.7 & $ < 0.21$ &  $< -4.97$ & $ -5.23$\tablenotemark{a} &        $<  0.2$ \\
            GJ~1005A & M4.0 &  3360 & $\le$3     &  0.8 & $ < 0.21$ &  $< -4.95$ & $ -5.05$\tablenotemark{c} &        $<  0.1$ \\
              GJ~299 & M4.5 &  3300 & $\le$3     &  0.8 & $ < 0.33$ &  $< -4.84$ & $<-5.55$\tablenotemark{a} &   $ 0.5\pm 0.2$ \\
             GJ~1227 & M4.5 &  3300 & $\le$3     &  1.0 & $ < 0.32$ &  $< -4.85$ & $<-3.86$\tablenotemark{a} &        $<  0.1$ \\
             GJ~1224 & M4.5 &  3300 & $\le$3     &  1.4 & $   4.13$ &  $  -3.74$ & $ -3.06$\tablenotemark{a} &   $ 2.7\pm 0.1$ \\
              YZ~Cmi & M4.5 &  3300 &    5       &  1.2 & $   8.42$ &  $  -3.43$ & $ -3.02$\tablenotemark{a} &        $>  3.9$ \\
              Gl~905 & M5.0 &  3240 & $\le$3     &  1.2 & $ < 0.30$ &  $< -4.95$ & $ -3.75$\tablenotemark{a} &        $<  0.1$ \\
             GJ~1057 & M5.0 &  3240 & $\le$3     &  1.2 & $   1.06$ &  $  -4.41$ & $<-3.87$\tablenotemark{a} &        $<  0.2$ \\
            GJ~1245B & M5.5 &  3150 &    7       &  2.1 & $   4.26$ &  $  -3.76$ & $ -3.58$\tablenotemark{a} &   $ 1.7\pm 0.2$ \\
             GJ~1286 & M5.5 &  3150 & $\le$3     &  1.7 & $   1.50$ &  $  -4.21$ & $<-3.77$\tablenotemark{a} &   $ 0.4\pm 0.2$ \\
             GJ~1002 & M5.5 &  3150 & $\le$3     &  1.5 & $ < 0.36$ &  $< -4.83$ & $<-5.24$\tablenotemark{a} &        $   0.0$ \\
              Gl~406 & M5.5 &  3150 &    3       &  2.6 & $   9.93$ &  $  -3.39$ & $ -2.77$\tablenotemark{a} &   $ 2.4\pm 0.1$ \\
             GJ~1111 & M6.0 &  2840 &   13       &  2.8 & $   8.12$ &  $  -3.92$ & $ -3.88$\tablenotemark{a} &   $ 1.7\pm 0.2$ \\
                VB~8 & M7.0 &  2620 &    5       &  3.4 & $   7.09$ &  $  -4.27$ & $ -3.47$\tablenotemark{d} &   $ 2.3\pm 0.2$ \\
            LHS~3003 & M7.0 &  2620 &    6       &  3.3 & $   7.60$ &  $  -4.24$ & $ -4.01$\tablenotemark{b} &   $ 1.5\pm 0.2$ \\
            LHS~2645 & M7.5 &  2540 &    8       &  3.1 & $   4.26$ &  $  -4.67$ &                           &   $ 2.1\pm 0.2$ \\
\tablebreak                                                        
         LP~412$-$31 & M8.0 &  2480 &    9       &  4.0 & $  23.05$ &  $  -3.89$ &                           &        $>  3.9$ \\
               VB~10 & M8.0 &  2480 &    6       &  4.2 & $   6.53$ &  $  -4.44$ & $ -4.9 $\tablenotemark{e} &   $ 1.3\pm 0.2$ \\
            LHS~2924 & M9.0 &  2390 &   10       &  4.3 & $   5.76$ &  $  -4.70$ & $ -4.35$\tablenotemark{d} &   $ 1.6\pm 0.2$ \\
            LHS~2065 & M9.0 &  2390 &   12       &  4.6 & $  29.05$ &  $  -4.00$ & $ -3.50$\tablenotemark{d} &        $>  3.9$ \\
  \enddata
\tablenotetext{a}{\cite{Delfosse98}}
\tablenotetext{b}{\cite{Fleming95}}
\tablenotetext{c}{$L_{\mathrm{X}}$ from \cite{NEXXUS}, $L_{\mathrm{bol}}$
    from the calibration of \cite{Delfosse98} with M$_{V}$ and $R-I$ from
    \citet{Gliese91}}
\tablenotetext{d}{\cite{Fleming93}}
\tablenotetext{e}{\cite{Fleming03}}
\end{deluxetable}

\clearpage
\thispagestyle{empty}
\clearpage

\begin{figure}
  \plotone{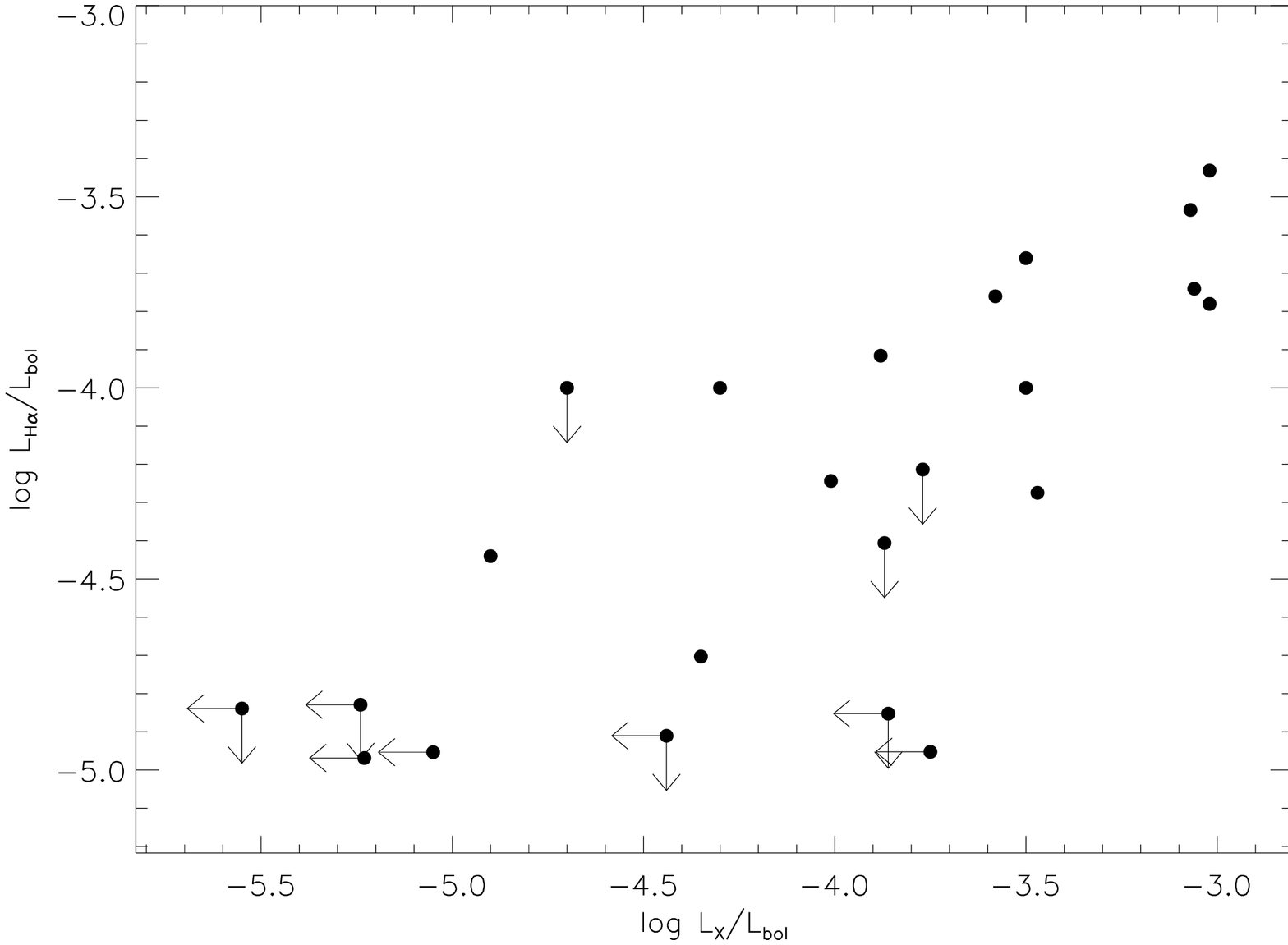}
  \caption{\label{fig:LalphaLx}$L_{\alpha}/L_{\mathrm{bol}}$ vs. $L_{\mathrm{X}}/L_{\mathrm{bol}}$}
\end{figure}

\end{document}